\documentclass[letterpaper,floatfix,aps,prb,amsmath,amssymb,twocolumn,
showpacs, nofootinbib]{revtex4}

\usepackage{epsfig}
\usepackage{verbatim}
\usepackage{epsf}
\usepackage{array}
\usepackage{hyperref}

\newcommand{\be}{\begin{equation}}
\newcommand{\ee}{\end{equation}}
\newcommand{\bea}{\begin{eqnarray}}
\newcommand{\eea}{\end{eqnarray}}

\renewcommand{\Re}{\mathrm{Re}\,}

\newcommand{\Tr}{\mathrm{Tr}}
\newcommand{\eref}[1]{Eq.~(\ref{#1})}
\newcommand{\esref}[1]{Eqs.~(\ref{#1})}
\newcommand{\rref}[1]{(\ref{#1})}

\newcommand{\ket}[1]{|#1\rangle}
\newcommand{\bra}[1]{\langle#1|}

\newcommand{\ocite}[1]{Ref.~\onlinecite{#1}}

\newcommand{\qp}{\mathrm{qp}}

\newcommand{\EZO}{\omega_{10}}

\begin{document}

\title{Decoherence of superconducting qubits caused by quasiparticle tunneling}

\author{G. Catelani}
\author{S. E. Nigg}
\author{S. M. Girvin}
\author{R. J. Schoelkopf}
\author{L. I. Glazman}
\affiliation{Departments of Physics and Applied Physics,
Yale University, New Haven, CT 06520, USA}

\begin{abstract}
In superconducting qubits, the interaction of the qubit degree of
freedom with quasiparticles defines a fundamental limitation for
the qubit coherence. We develop a theory of the pure dephasing rate
$\Gamma_\phi$ caused by quasiparticles tunneling through a Josephson junction
and of the inhomogeneous broadening due to changes in the occupations of
Andreev states in the junction.
To estimate $\Gamma_\phi$, we derive a master equation for the qubit
dynamics. The tunneling rate of free quasiparticles is enhanced by their
large density of states at energies close to the superconducting
gap. Nevertheless, we find that $\Gamma_\phi$ is small compared to the
rates determined by extrinsic factors in most of the current qubit
designs (phase and flux qubits, transmon,
fluxonium). The split transmon, in which a single junction is replaced by a SQUID loop,
represents an exception that could
make possible the measurement of $\Gamma_\phi$. Fluctuations of the
qubit frequency leading to
inhomogeneous broadening may be caused by the
fluctuations in the occupation numbers of the Andreev states associated
with a phase-biased Josephson junction. This mechanism
may be revealed in qubits with small-area junctions,
since the smallest relative change in frequency it causes is of the order
of the inverse number of transmission channels in the junction.
\end{abstract}

\date{\today}


\maketitle

\section{Introduction}

Over the past several years significant efforts have been directed toward designing and implementing
superconducting circuits with improved coherence properties. For quantum computation purposes, the
coherence time $T_2$ of a qubit must be sufficiently long as to allow for error correction.\cite{div}
The unavoidable couplings of the qubit with various sources of noise are responsible for decoherence,
and different types of qubits have different sensitivities to a given noise source.
For example, the phase and flux qubits coherence times are limited by flux noise,\cite{Bialczak,Yoshihara}
while the transmon parameters are chosen to decrease the effect of charge noise in comparison with the Cooper
pair box.\cite{transmon} Flux and charge noise originate from the environment surrounding the qubits;
in this paper, by contrast, we study an \textit{intrinsic} mechanism of decoherence due to
the coupling between the qubit and the quasiparticle excitations in the superconductor the qubit is made of.
In general one can distinguish two contributions to the time $T_2$: first, the qubit can
lose energy and the corresponding relaxation time $T_1$ imposes an upper bound to
the coherence time, $T_2 \le 2T_1$. Second, additional
pure dephasing mechanisms, characterized by the rate $\Gamma_\phi$, can shorten $T_2$ below this upper limit.
Recent theoretical\cite{prl,prb} and experimental\cite{paik, martinis,sun}
works have highlighted the contribution of quasiparticle tunneling to the relaxation rate.
Here we focus on the pure dephasing effect of quasiparticle
tunneling.

The decoherence rates discussed above are related to the power spectral density $S(\omega)$ of the noise source:
the relaxation rate is proportional to the value of the spectral density at the qubit frequency $\EZO$,
$1/T_1 \propto S(\EZO)$, while the pure dephasing rate is determined by the low-frequency part of the
spectral density, $\Gamma_\phi \sim S(0)$ -- see, e.g., \ocite{ithier}.
Clearly the latter relationship cannot hold if the power spectral
density diverges as $\omega \to 0$. Because of its experimental relevance, a well-studied example of
diverging spectral density is that of $1/f$ noise; in the case of $1/f$ flux noise, for instance,
the decay of the qubit coherence is not exponential in time, but Gaussian-like\cite{ithier,mart_deph}
(up to a logarithmic factor that depends on the measurement protocol).
In studying how quasiparticle tunneling affects dephasing we find another such example,
since the quasiparticle current
spectral density is logarithmically divergent at low frequencies when the gaps on
the two sides of the junction have the same magnitudes (see Sec.~\ref{sec:me}).
We
show that despite this divergence, a finite dephasing rate $\Gamma_\phi$
can be determined. We then estimate the dephasing rate for a few different single- and multi-junction qubits
and find that in most cases $\Gamma_\phi$ is small compared to the the quasiparticle induced relaxation
rate. An exception is the split transmon, in which the two rates can be of the same order
of magnitude (see Sec.~\ref{sec:st}). Since it is known that quasiparticles limit
the relaxation rate in this system at sufficiently high temperatures,\cite{sun}
it may be possible to measure the quasiparticle dephasing rate if other sources of dephasing can be minimized.

The quasiparticle dephasing mechanism discussed above is due to tunneling of \textit{free} quasiparticles
across the junction. Another dephasing mechanism originates from quasiparticles weakly \textit{bound}
to a phase-biased junction that give rise to subgap Andreev states; the dephasing is caused by changes in the
occupations of these states that make the Josephson coupling and hence qubit frequency $\omega_q$ fluctuate.
Because of this additional dephasing, the measured decoherence rate $1/T_2^*$ acquires an
inhomogeneous broadening contribution, $1/T_2^*-1/T_2$, which can be suppressed using echo pulse
sequences.
When the average occupation $x_\qp^A$ of the Andreev states is small, $x_\qp^A\ll 1$, the typical
(i.e., root mean square) fluctuation of the occupations is given by the square
root of $x_\qp^A$. Then for the phase qubit we show in Sec.~\ref{sec:and}
that the typical frequency fluctuation is proportional to the typical
fluctuation of the occupations divided by the square root of the (effective) number
of transmission channels $N_e$ in the junction,
$\langle (\Delta\omega_q)^2\rangle^{1/2}/\omega_q \propto \sqrt{x_\qp^A/N_e}$. For these fluctuations
to measurably affect the decoherence rate $1/T_2^*$,
the condition $\langle{\Delta\omega_q}^2\rangle^{1/2}T_2 \gtrsim 1$ should be satisfied;
using this condition we estimate that this mechanism
is not a limiting factor to coherence in current experiments with phase qubits.
On the contrary, it could contribute to decoherence in recent transmon experiments,\cite{paik,sears}
due to the small junction area (i.e., smaller $N_e$ in comparison with phase qubits).
However, this possibility will require a separate investigation, due to the lack of phase
bias in the transmon.

The paper is organized as follows: in the next Section we introduce the effective description
of a single-junction system. In Sec.~\ref{sec:me} we present the master
equation governing the qubit dynamics and we discuss the self-consistent regularization
of the logarithmic divergence in the dephasing rate. Applications of our results to single- and multi-junctions
qubits are in Secs.~\ref{sec:single} and \ref{sec:multi}, respectively.
The role of Andreev states is analyzed in Sec.~\ref{sec:and}.
We summarize our work in
Sec.~\ref{sec:summ}.
We use units $\hbar = k_B = 1$ throughout the paper.

\section{Effective model}
\label{sec:model}

The effective Hamiltonian $\hat{H}$ for a superconducting qubit can be
split into two parts,
\be\label{Heff}
\hat{H} = \hat{H}_0 + \delta\hat{H}\, ,
\ee
where the non-interacting Hamiltonian $\hat{H}_0$ is the sum of qubit and quasiparticle terms,
\be
\hat{H}_0 = \hat{H}_\varphi +\hat{H}_\qp\, .
\ee
The Hamiltonian for the qubit degree of freedom accounts for the charging ($E_C$), Josephson ($E_J$),
and inductive ($E_L$) energies in a system comprising an inductive loop shunting a tunnel junction,
\be\label{Hq0}
\hat{H}_\varphi = 4E_C \left(\hat{N}-n_g\right)^2 -E_J \cos \hat\varphi
+ \frac12 E_L\!\left(\hat\varphi-2\pi\Phi_e/\Phi_0\right)^2,
\ee
with $n_g$ the dimensionless gate voltage, $\Phi_e$ the external magnetic flux
threading the loop, and $\Phi_0=h/2e$ the flux quantum. The operator $\hat{N}=-id/d\varphi$
counts the number of Cooper pairs passed through the junction.
The quasiparticle Hamiltonian is given by
\be\label{Hqp}
\hat{H}_\qp =\sum_{j=L,R} \hat{H}_{\qp}^j \, , \quad \hat{H}_{\qp}^j = \sum_{l=1}^{N_{ch}}
\sum_{n,\sigma} \epsilon_{n}^j
\hat\alpha^{j\dagger}_{n\sigma l} \hat\alpha^j_{n\sigma l},
\ee
where $\hat\alpha^j_{n\sigma l}$($\hat\alpha^{j\dagger}_{n\sigma l}$) are annihilation (creation)
operators for quasiparticles with channel index $l$ and spin $\sigma=\uparrow, \downarrow$ in lead $j=L,R$ to
the left or right of the junction.
We have assumed for simplicity the same number of channels $N_{ch}$ and identical
densities of states per spin direction $\nu_0$ in both leads.
Denoting with $\Delta^j$ the superconducting gap, the
quasiparticle energies are $\epsilon^j_{n} =
\sqrt{(\xi_{n}^j)^2+(\Delta^j)^2}$, with $\xi_{n}^j$
single-particle energy level $n$ in the normal state of lead $j$.
The occupation probabilities of these levels are given by the distribution functions
\be
f^j(\xi^j_{n})= \langle\!\langle \hat\alpha_{n\uparrow l}^{j\dagger} \hat\alpha^j_{n\uparrow l}
\rangle\!\rangle_\qp
= \langle\!\langle \hat\alpha_{n\downarrow l}^{j\dagger} \hat\alpha^j_{n\downarrow l}\rangle\!\rangle_\qp\, ,
\quad j=L,R\, ,
\ee
where double angular brackets
$\langle\!\langle \ldots \rangle\!\rangle_\qp$ denote averaging over quasiparticle states.
We take the distribution functions to be independent of spin and equal in the two leads. We also
assume that $\delta E$, the characteristic energy of the quasiparticles above the gap, is small compared
to the gap, but
the distribution function is otherwise generic, thus allowing for non-equilibrium conditions.

The interaction term $\delta\hat{H}$ in \eref{Heff} accounts for tunneling and, as discussed in Appendix~A
of \ocite{prb}, is the sum of three parts: quasiparticle tunneling $\hat{H}_T$, pair tunneling $\hat{H}_T^p$,
and the Josephson energy counterterm $\hat{H}_{E_J}$. When the superconducting gaps are larger than all other
energy scales, the only effect of the last two terms is to contribute to the renormalization of the qubit
frequency\cite{prb} [see also the discussion after \eref{Asf_def}];
therefore, we neglect those terms and consider only the quasiparticle tunneling Hamiltonian,
$\delta\hat{H} = \hat{H}_T$ with
\bea
\hat{H}_T &=& \sum_{l,k=1}^{N_{ch}}\!\!\tilde{t}_{lk}\!\!\sum_{n,m,\sigma}\!\!
\left(e^{i\hat\varphi/2} u_{n}^L u_{m}^R
- e^{-i\hat\varphi/2} v_{m}^R v_{n}^L
\right)\hat\alpha_{n\sigma l}^{L\dagger} \hat\alpha^R_{m\sigma k}
\nonumber \\ \label{HT0} &+& \text{H.c.}\quad
\eea
Here the Bogoliubov amplitudes $u^j_{n}$, $v^j_{n}$ are real quantities, since their dependence
on the phases of the order parameters appears explicitly through the gauge-invariant phase difference $\varphi$.
The elements $\tilde{t}_{lk} \ll 1$ of the electron tunneling matrix $\tilde{t}$ are related to the junction
conductance by $g_T = 2 g_K \sum_{p=1}^{N_ch} \mathrm{T}_p$, where $g_K=e^2/h$ is the conductance quantum and
the transmission probabilities $\mathrm{T}_p$ ($p=1,\ldots,N_{ch}$) are
the eigenvalues of the matrix $(2\pi \nu_0)^2 \tilde{t}\tilde{t}^{\dagger}$.

Since we are interested in the dynamics of the qubit only, rather than that of a multi-level system,
we project the Hamiltonian $\hat{H}$ onto the qubit states $|0\rangle$ and $|1\rangle$,
which we represent by the vectors $(0, 1)^T$ and $(1,0)^T$ for the ground and excited states,
respectively; the two-level approximation is justified under the conditions
that permit the operability of the system as a qubit\cite{DevMar} (i.e., anharmonicity large
compared to linewidth). Then in terms of the Pauli matrices we can write
\be\label{Hq1}
\hat{H}_\varphi = \frac{\EZO}{2} \hat\sigma^z \, ,
\ee
where the qubit frequency in general depends on all the parameters present in \eref{Hq0}, and,
dropping for notational simplicity the channel indices,\cite{ch_ind}
\be\label{HT}\begin{split}
\hat{H}_T = \tilde{t} \sum_{n,m,\sigma}
\Big[A_{nm}^{d} \hat\sigma^z + A_{nm}^{r} \left(\hat\sigma^+ + \hat\sigma^- \right)
\\ + A_{nm}^{f} \hat{I} \Big]
\hat\alpha^{L\dagger}_{n\sigma} \hat\alpha^{R}_{m\sigma}
+ \text{H.c.} \, ,
\end{split}\ee
where the coefficients $A_{nm}^{k}$, $k=d,\, r,\, f$, have the structure
\be
A_{nm}^{k} = A_c^k \left(u^L_n u^R_m - v^L_n v^R_m \right) +
i A_s^k \left(u^L_n u^R_m + v^L_n v^R_m\right) \, .
\ee
Here $A_{c,s}^k$ denote combinations of matrix elements for the operators $e^{\pm i\hat\varphi/2}$
associated with the transfer of a single charge across the junction,
\bea\label{medef}
s_{ij} &=& \langle i| \sin\frac{\hat\varphi}{2} |j \rangle \\
\label{Asdef}
A_s^d & = & \frac12\left(s_{11}
- s_{00} \right)
\\ \label{Asr_def}
A_s^r & = & s_{10}
\\
\label{Asf_def}
A_s^f & = & \frac12\left(s_{11}
+ s_{00} \right)
\eea
and the $A_{c}^k$ are obtained by replacing sine with cosine in the above definitions. As it will become evident
in the next section, only the terms with $k=d$ and $k=r$ contribute to pure dephasing and relaxation of the qubit,
respectively.

The term with $k=f$ (in combination with the $k=r$ one) contributes to the average frequency shift.
More precisely, the average frequency shift $\delta\omega = \delta\omega_{E_J} + \delta\omega_\qp$ has two
parts,\cite{prb} originating from the quasiparticle renormalization of the Josephson energy and virtual
transitions between qubit states mediated by quasiparticles, respectively.
The latter part ($\delta \omega_\qp$) is discussed further in Appendix~\ref{app:me}. Here we note
that in the
leading ($\propto \tilde{t}^2$) order, the Josephson part $\delta\omega_{E_J}$ is
the sum of two contributions with distinct origins.
The first one comes from the product of the terms proportional to $A_{nm}^{f}$ and
$A_{nm}^{r}$ in $\delta\hat{H}_T$ [\eref{HT}].
The second contribution is due to
the terms we neglected in $\delta\hat{H}$. (The neglected terms are the pair tunneling and Josephson counterterm,
as defined in Appendix~A of \ocite{prb}.)
Since we are studying decoherence effects in this work,
we set $A_{nm}^{f}=0$
henceforth. Equations \rref{Hqp}, \rref{Hq1}, and \rref{HT} (with $A_{nm}^{f}=0$) constitute the
starting point for the derivation of the master equation presented in the next section.

\section{Qubit phase relaxation: the master equation}
\label{sec:me}

The information on the time evolution of the qubit is contained in its density matrix $\hat\rho(t)$,
which we decompose as
\be
\hat\rho = \frac12 \left[\hat{I} + \rho_z \hat\sigma^z\right]
+ \rho_+ \hat\sigma^- + \rho_+^* \hat\sigma^+
\ee
In this section we present the final form of the master equation for the density matrix.
The derivation can be found
in Appendix~\ref{app:me}, where we
start from the Hamiltonian of the system presented in the previous section
and employ the standard Born-Markov and secular (rotating wave) approximations \cite{petru}
to arrive at the expressions given here.

The diagonal component $\rho_z$ of the density matrix obeys the equation
\be\label{rz_me}
\frac{d \rho_z}{dt} = -\left[\Gamma_{1\to 0} + \Gamma_{0\to 1}\right] \rho_z
+ \left[\Gamma_{0\to 1} - \Gamma_{1\to 0}\right]
\ee
where, assuming equal gaps in the leads ($\Delta^L = \Delta^R \equiv \Delta$),
\be\label{G10}\begin{split}
\Gamma_{1 \to 0} = & \frac{2 g_T}{\pi g_K} \int_\Delta^{+\infty}\!\!d\epsilon
 \, f(\epsilon)\left(1-f(\epsilon+\EZO)\right) \\
\bigg[ & \frac{\epsilon(\epsilon+\EZO)+\Delta^2}{\sqrt{\epsilon^2-\Delta^2}
\sqrt{(\epsilon+\EZO)^2-\Delta^2}}\left|A^r_s\right|^2
\\ + & \frac{\epsilon(\epsilon+\EZO)-\Delta^2}{\sqrt{\epsilon^2-\Delta^2}
\sqrt{(\epsilon+\EZO)^2-\Delta^2}}\left|A^r_c\right|^2
\bigg]
\end{split}\ee
and $\Gamma_{0\to 1}$ is obtained by the replacement $f \to 1-f$. Here $g_K = e^2/h$ is
the conductance quantum. The general solution to \eref{rz_me}
is
\be
\rho_z (t) = \rho_z(0) e^{-t/T_1} +
\frac{\Gamma_{0\to 1} - \Gamma_{1\to 0}}{\Gamma_{0\to 1} + \Gamma_{1\to 0}}
\ee
where we introduced the relaxation time $T_1$ as
\be\label{t1def}
\frac{1}{T_1} = \Gamma_{0\to 1} + \Gamma_{1\to 0} \, .
\ee

Equation \rref{G10} represents the generalization, valid for any $\EZO < 2\Delta$, of the relaxation
rate formula derived in Refs.~\onlinecite{prl,prb} in the limit $\EZO \ll 2\Delta$ using Fermi's golden rule.
Indeed, the assumption that quasiparticles have characteristic energies small compared to the gap
enables us to approximately substitute $\epsilon \to \Delta$ in the numerators in square brackets in
\eref{G10}, and neglecting terms of order $\EZO/\Delta$ we find
\be\label{G10l}
\Gamma_{1\to 0} \simeq \left|A_s^r\right|^2 S_\qp(\EZO)\, ,
\ee
where
\be\label{Sqp}\begin{split}
S_\qp(\omega) = \frac{16E_J}{\pi} \int_0^{+\infty}\!\!dx \, \frac{1}{\sqrt{x}\sqrt{x+\omega/\Delta}}
 \\ f\left[(1+x)\Delta\right]\left\{1-f\left[(1+x)\Delta+\omega\right]\right\}
\end{split}\ee
and we remind that $E_J = \Delta g_T/8g_K$.
The agreement of \eref{G10l} with the results of Refs.~\onlinecite{prl,prb} validates the present
approach. Since the relaxation rate is studied in detail in those references, we do not consider it
here any further, except to
note that the terms neglected in \eref{G10l} can become important if the matrix element $A_s^r$ is small,
$|A_s^r/A_c^r|^2 \lesssim \EZO/\Delta$.
In fact, $A_s^r$ can vanish at particular values
of the external parameters used to tune the qubit, for example in the flux qubit when the external flux
equals half the flux quantum;\cite{prl,prb} in such a case, one needs to retain
the term proportional to $A_c^r$ in \eref{G10} to evaluate the (non-vanishing) relaxation rate.

The master equation for the off-diagonal part of the density matrix is
\be\label{rp_me}
\frac{d \rho_+}{dt} = i\left(\EZO + \delta\omega \right)\rho_+ - \frac{1}{2T_1} \rho_+
- \Gamma_\phi \rho_+
\ee
where $\delta\omega$ is the quasiparticle-induced average frequency shift\cite{prl,prb} discussed in the previous
section,
$T_1$ is defined in \eref{t1def}, and the pure dephasing rate is
\be\label{Gp_def}\begin{split}
\Gamma_\phi = & \frac{4 g_T}{\pi g_K}\int_{\Delta^R}^{+\infty}\!\!d\epsilon
 \, f(\epsilon)\left[1-f(\epsilon)\right] \\
\bigg[ & \frac{\epsilon^2+\Delta^L \Delta^R}{\sqrt{\epsilon^2-(\Delta^L)^2}
\sqrt{\epsilon^2-(\Delta^R)^2}}\left|A^d_s\right|^2
\\ + & \frac{\epsilon^2-\Delta^L\Delta^R}{\sqrt{\epsilon^2-(\Delta^L)^2}
\sqrt{\epsilon^2-(\Delta^R)^2}}\left|A^d_c\right|^2
\bigg]
\end{split}\ee
where we assumed $\Delta^R > \Delta^L$. The general solution to \eref{rp_me} is
\be\label{p_t}
\rho_+(t) = \rho_+(0) e^{i(\EZO+\delta\omega)t} e^{-t/T_2}
\ee
with
\be\label{t2def}
\frac{1}{T_2} = \frac{1}{2T_1} + \Gamma_\phi \, .
\ee

The pure dephasing rate defined in \eref{Gp_def} has a structure similar to that of the
relaxation rate, \eref{G10}, if we substitute $\EZO \to 0$ and $A^r_{s(c)} \to A^d_{s(c)}$ in the latter.
Thus we recover the
relationship between the power spectral density $S(\omega)$ of a noise source
and the decoherence rates discussed in the Introduction,
$\Gamma_{1 \to 0} \propto S(\EZO)$ and $\Gamma_\phi \propto S(0)$.
However, in \eref{Gp_def} we have explicitly assumed an asymmetric junction, $\Delta^R > \Delta^L$,
and extension of this result to the typical case of a symmetric junction
($\Delta^R = \Delta^L$) is problematic. Indeed, let us consider an almost symmetric junction,
$\Delta^R - \Delta^L \ll \Delta^R$, with $|A^d_s| \gtrsim |A^d_c|$ and a non-degenerate
quasiparticle distribution [$f(\epsilon) \ll 1$, $\epsilon > \Delta^R$]; then we find,
using from now on the notation $\Delta = \Delta^R$,
\be\label{Gp_l}\begin{split}
\Gamma_\phi & \simeq \frac{4g_T}{\pi g_K} \left|A^d_s\right|^2 \Delta
\int_0^{+\infty}\!dx \, \frac{f\left[(1+x)\Delta \right]}
{\sqrt{x}\sqrt{x+\left(\Delta-\Delta^L\right)/\Delta}}
\\ & \simeq 2 \left|A^d_s\right|^2 S_\qp\left(\Delta-\Delta^L\right)
\end{split}\ee
In the symmetric junction limit $\Delta^L \to \Delta$, $\Gamma_\phi$ diverges logarithmically
due to the singularity at $x=0$ of the integrand in \eref{Gp_l}; for example, in thermal equilibrium
at temperature $T \gg \Delta- \Delta^L$ we have
\be
\Gamma_\phi \approx \frac{32 E_J}{\pi} \left|A^d_s\right|^2 e^{-\Delta/T} \left[
\ln \frac{4T}{\Delta-\Delta^L} - \gamma_E \right]
\ee
Due to the logarithmic divergence, in general we cannot simply take $\Gamma_\phi \propto S_\qp(0)$;
the correct procedure that leads to a finite dephasing rate is presented in the next section.

\subsection{Self-consistent dephasing rate}
\label{sec:deph}
The terms in the right hand sides of the master equations \rref{rz_me} and \rref{rp_me}
are proportional to the square of the tunneling amplitude via the
tunneling conductance $g_T \propto \tilde{t}^2$;
this proportionality is a consequence of the lowest order perturbative treatment of the
tunneling Hamiltonian [\eref{HT}], which enables us to neglect higher order (in $\tilde{t}$) terms
when evaluating certain correlation functions involving qubit and quasiparticle operators
[see Appendix~\ref{app:me} for details]. This implies that those correlation functions oscillate but
do not decay in time, which is a limitation of the used approximation:
the inclusion of higher order effects introduces decaying factors
of the from $e^{-\gamma t}$ into the correlation functions, where at leading order the
decay rate $\gamma$ is itself proportional to the tunneling conductance.
Here we discuss an Ansatz for $\gamma$ whose validity is checked perturbatively in Appendix \ref{app:scd}.
As we show there, a finite decay rate $\gamma$ reflects itself into a smearing of
the singularity for $\Delta^L = \Delta$ of the integrand in \eref{Gp_l},
\be\label{smearing}\begin{split}
\int_0^{+\infty} \frac{dx}{x} = & \int_0^{+\infty} \frac{dx}{\sqrt{x}} \int_0^{+\infty}
\frac{dy}{\sqrt{y}} \delta(x-y) \ \to \\ & \int_0^{+\infty} \frac{dx}{\sqrt{x}}
\int_0^{+\infty} \frac{dy}{\sqrt{y}} \frac{1}{\pi}\frac{\gamma/\Delta}{(x-y)^2 + (\gamma/\Delta)^2}
\end{split}\ee

In the problem at hand there are
two
inverse time scales which could serve as a low-energy
cut-off to regularize the integral as in the above equation,
the relaxation rate $\Gamma_{1 \to 0}$ and the pure dephasing rate $\Gamma_\phi$.
A finite relaxation rate means that the qubit excited level has a finite width; one could argue that
this uncertainty in the energy will in turn reflect itself in an uncertainty of the energy exchanged between
qubit and quasiparticles, thus smearing the singularity as in \eref{smearing}.
However,
relaxation rate and dephasing rate are
determined by different matrix elements [cf. \esref{Asdef}-\rref{Asr_def}], so one can imagine,
at least in principle, a limiting situation in which the relaxation rate vanishes, which would then cause
the dephasing rate to diverge.
Therefore,
we expect that dephasing processes
will themselves be the ultimate limiting factors for coherence, so that $\gamma = \Gamma_\phi$.
With this identification, we arrive at the self-consistent expression for the pure
dephasing rate
\be\label{Gp_sc}\begin{split}
\Gamma_\phi = \frac{32 E_J}{\pi} \left|A^d_s\right|^2 \int_0^{+\infty}\!\frac{dx}{\sqrt{x}}
\int_0^{+\infty}\!\frac{dy}{\sqrt{y}}
f\left[(1+x)\Delta\right] \\ \times \left\{1-f\left[(1+y)\Delta\right]\right\} \frac{1}{\pi}
\frac{\Gamma_\phi/\Delta}{\left(x-y\right)^2+\left(\Gamma_\phi/\Delta\right)^2}
\end{split}\ee
Equation \rref{Gp_sc} is the central result of this paper. It is valid for symmetric junctions
(or nearly symmetric, $\Delta^R - \Delta^L \ll \Gamma_\phi$) and we show in Appendix~\ref{app:scd} that
it agrees with the result of the perturbative derivation of the master equation
extended with logarithmic accuracy to the next to leading order in $\tilde{t}^2$.

Similarly to the relaxation rate, for some specific values of the qubit parameters
the matrix element $A^d_s$ can be small or even vanish exactly. Then one should take into account the
second term in square brackets in \eref{Gp_def} to get
\be\label{Gp_sl}
\Gamma_\phi = \frac{32 E_J}{\pi} \left|A^d_c\right|^2 \int_0^{+\infty}\!\!dx \,
f\left[(1+x)\Delta\right] \left\{1-f\left[(1+x)\Delta\right]\right\}
\ee
An estimate for the actual dephasing rate is given by the larger of the two rates calculated using
\eref{Gp_sc} or \eref{Gp_sl}.

\subsection{Non-equilibrium quasiparticles}
\label{sec:nebr}

The relaxation rate in \eref{G10l}
depends explicitly on the qubit
properties via the matrix element $A_s^r$,
while the spectral density $S_\qp$
accounts for the dynamics of quasiparticle tunneling.
The same structure is present in the right hand sides of \esref{Gp_sc}-\rref{Gp_sl} --
a matrix element multiplies factors describing the tunneling dynamics.
These factors can be further
simplified under certain assumptions.
Here we focus on \eref{Gp_sc} and distinguish two cases: first, let us assume that the quasiparticle
energy is small compared to the dephasing rate, $\delta E \ll \Gamma_\phi$, and that quasiparticles are
non-degenerate, $f[(1+y)\Delta]\ll 1$. Then integrating first over $y$ and then over $x$ we find
\be\label{Gp_des}
\Gamma_\phi \simeq \frac{16E_J}{\pi} \left|A^d_s\right|^2 \sqrt{\frac{\Delta}{\Gamma_\phi}} \, x_\qp \, ,
\ee
where
\be\label{xqp_def}
x_\qp = \sqrt{2}\int_0^{+\infty}\!\frac{dx}{\sqrt{x}} \, f\left[(1+x)\Delta\right]
\ee
is the quasiparticle density normalized by the density of Cooper pairs. Indicating with $f_0$ the typical
occupation probability, we estimate\cite{f0est}
$x_\qp \sim f_0 \sqrt{\delta E/\Delta}$.
Then solving \eref{Gp_des} for $\Gamma_\phi$, the requirement $\Gamma_\phi \gg \delta E$ can be written as
\be
\frac{16}{\pi} \frac{E_J}{\Delta} \left|A^d_s\right|^2 f_0 \gg \frac{\delta E}{\Delta}
\ee
This condition is in practice difficult to satisfy, since with our assumptions $f_0 \ll 1$, while
$\left|A^d_s\right| \leq 1$, $E_J/\Delta \lesssim 1$, and at the lowest experimental temperatures
$\delta E/\Delta \sim T/\Delta \gtrsim 0.01$. Thus we conclude that for non-degenerate quasiparticles
an upper bound for the dephasing rate is given by $\Gamma_\phi \lesssim \delta E$.

The second case we consider, for both degenerate and non-degenerate quasiparticles, is in fact that of small
dephasing rate, $\Gamma_\phi \ll \delta E$. Then neglecting terms of order $\Gamma_\phi/\delta E$,
\eref{Gp_sc} simplifies to
\be\label{Gp_scs}\begin{split}
\Gamma_\phi \simeq  & \ \frac{32 E_J}{\pi} \left|A^d_s\right|^2 \int_0^{+\infty}\!\frac{dx}{\sqrt{x}} \Re
\frac{1}{\sqrt{x+i\Gamma_\phi/\Delta}} \\ & \times
f\left[(1+x)\Delta\right] \left\{1-f\left[(1+x)\Delta\right]\right\} \\
\sim & \ \frac{32 E_J}{\pi} \left|A^d_s\right|^2 f_0 \left(1-f_0\right) \, \ln \frac{4\delta E}{\Gamma_\phi}
\end{split}\ee
We note that both \esref{Gp_des} and \rref{Gp_scs} can be written approximately in the form\cite{shnirman}
$\Gamma_\phi \propto \left|A^d_s\right|^2 S_\qp (\Gamma_\phi)$; however, the proportionality coefficients
are different in the two cases.
Solving \eref{Gp_scs} for $\Gamma_\phi$ by iterations gives
\be
\Gamma_\phi \approx \frac{32 E_J}{\pi} \left|A^d_s\right|^2 f_0 \left(1-f_0\right) \,
\ln \frac{\pi\delta E}{8 E_J \left|A^d_s\right|^2 f_0 \left(1-f_0\right)} \, .
\ee
As a specific example, we consider from now on a quasi-equilibrium distribution
$f(\epsilon) = e^{-\epsilon/T_e}$,
where $T_e$ is the effective quasiparticle temperature.\cite{note_te}
In this case we have $\delta E = T_e$ and $f_0 = e^{-\Delta/T_e} \ll 1$, so that the
dephasing rate is
\be\label{Gp_Te}
\Gamma_\phi(T_e) \approx \frac{32 E_J}{\pi} \left|A^d_s\right|^2 e^{-\Delta/T_e}
\left[\frac{\Delta}{T_e} + \ln \frac{\pi T_e}{8 E_J \left|A^d_s\right|^2} \right].
\ee

To conclude this section, we note that the divergence for $\Delta^R=\Delta^L$ in \eref{Gp_def}
is a consequence of the square root singularity of the BCS density of states
at the gap edge. Therefore possible modifications of the density of states (e.g., broadening\cite{dynes}) would
in principle lead to different estimates of the dephasing rate;
the effect of a small density of subgap states has been recently considered in \ocite{marth2}.
However, we argue in Appendix~\ref{app:broad}
that these potential modifications are not relevant to current experiments with Al-based qubits, which
we focus on for the remainder of the paper.

\section{Phase relaxation of single-junction qubits}
\label{sec:single}

In this section we consider the dephasing rate for two single-junction systems, the phase
qubit and the transmon, under the assumption of small qubit
frequency, $\EZO \ll \Delta$ (see Appendix~\ref{app:fl} for the flux qubit).
The calculations of the matrix element entering the relaxation rate
are described in detail in \ocite{prb}, whose result we briefly summarize.
Here we use (without giving all the details) the same approach of that work to obtain the matrix
elements for dephasing. Interestingly, in all cases the pure dephasing rate $\Gamma_\Phi$ turns out to add
at most a small correction to $1/T_2$ in comparison with the relaxation term $1/2T_1$.

\subsection{Phase qubit}
\label{sec:ph}

In a phase qubit, the charging energy $E_C$ is small compared to the transition frequency $\EZO$. The
latter depends on the external flux via the position $\varphi_0$ of a minimum in the potential energy
of the Hamiltonian in \eref{Hq0}, as determined by
\be\label{ph_min}
E_J \sin \varphi_0 + E_L \left(\varphi_0 - 2\pi \Phi_e/\Phi_0\right) = 0
\ee
Then the frequency is
\be\label{ph_fr}
\EZO = \sqrt{8E_C \left(E_L + E_J \cos \varphi_0 \right)} \, .
\ee
For a small effective temperature $T_e \ll \EZO$ the relaxation time is
\be\label{ph_rel}
\frac{1}{T_1} = \frac{1}{\pi}\frac{\omega_p^2}{\EZO} \, e^{-\Delta/T_e} \sqrt{\frac{\pi T_e}{\EZO}}
\left(1+\cos\varphi_0\right) \, ,
\ee
where
\be
\omega_p =\sqrt{8 E_C E_J}
\ee
is the plasma frequency of the junction.

Within the same approximations used to obtain the above formulas,\cite{phq_app}
the matrix element for dephasing is
\be\label{ph_me}
\left|A^d_s\right|^2 = \frac{1}{8} \left(\frac{E_C}{\EZO}\right)^2 \left(1-\cos\varphi_0\right)
\ee
and substituting into \eref{Gp_Te} we get
\be\label{ph_dep}\begin{split}
\Gamma_\phi = \frac{E_C}{2\pi} \frac{\omega_p^2}{\EZO^2}e^{-\Delta/T_e}
\left[\frac{\Delta}{T_e} + \ln \frac{8\pi T_e \EZO^2}{E_C \omega_p^2 (1-\cos\varphi_0)}\right]
\\ \times (1-\cos\varphi_0).
\end{split}\ee
Note that the factor in front of $e^{-\Delta/T_e}$ is smaller for $\Gamma_\phi$ in comparison with that
for $1/T_1$ because the matrix element for dephasing is smaller than that for relaxation by a factor
$E_C/\EZO$. At low temperatures the terms in square brackets
in \eref{ph_dep} are dominated by $\Delta/T_e$ and hence, neglecting factors $\cos\varphi_0$ as they are
small compared to unity,
the condition $2T_1\Gamma_\phi > 1$ can be written as
\be\label{ph_Tecond}
\frac{T_e}{\Delta} < \left(\frac{\EZO}{\Delta}\right)^{1/3} \left(\frac{E_C}{\EZO}\right)^{2/3}
\ee
Typically for a phase qubit the product on the right is of order $10^{-2}$, while
$T_e/\Delta \sim 10^{-1}$. Therefore the pure dephasing contribution
to $T_2$ [\eref{t2def}] can be neglected. Interestingly, for a quasiparticle temperature
of the order of the base temperature, $T/\Delta \sim 10^{-2}$, relaxation and pure dephasing
would have similar order of magnitudes, although both would be much smaller than at $T_e/\Delta \sim 10^{-1}$
due to their common exponential suppression by the Boltzmann factor.

\subsection{Transmon}

The Hamiltonian of  transmon is given by \eref{Hq0} with $E_L=0$,
supplemented by a periodic boundary condition in phase.\cite{transmon}
For our purposes, the transmon can be considered as a particular case of the phase qubit
with $\varphi_0=0$ [see \eref{ph_min}].
With these parameters, one obtains from \eref{ph_rel} the correct estimate for the relaxation time $T_1$,
\be\label{tr_rel}
\frac{1}{T_1} = \frac{2}{\pi} \omega_p e^{-\Delta/T_e} \sqrt{\frac{\pi T_e}{\omega_p}}
\ee
However, the vanishing for $\varphi_0=0$ of the matrix element in \eref{ph_me} is not the correct result for
the transmon: careful evaluation of the matrix element, following
the procedure outlined in Appendices B and C of \ocite{prb}, gives an exponentially small value,
$A^d_s \propto \exp\left[-\sqrt{8 E_J/E_C}\right]$. This exponential suppression is sufficient to
ensure that the dephasing rate is dominated by the contribution in \eref{Gp_sl}, since the matrix
element entering that equation has no such suppression,
\be
\left|A^d_c\right|^2 = \frac{1}{4} \left(\frac{E_C}{\omega_p}\right)^2 = \frac{1}{32}\frac{E_C}{E_J}
\ee
Substituting this expression into \eref{Gp_sl}, for the quasi-equilibrium distribution function we find
\be\label{tr_dep}
\Gamma_\phi = \frac{1}{\pi} E_C e^{-\Delta/T_e} \frac{T_e}{\Delta}
\ee
Using \esref{tr_rel} and \rref{tr_dep} it is easy to show that for the transmon
$2T_1 \Gamma_\phi \ll 1$; therefore, as for the phase qubit, the pure dephasing contribution to
$T_2$ is negligible.

\section{Phase relaxation of multi-junction qubits}
\label{sec:multi}

The results of Sec.~\ref{sec:me} are readily generalized to multi-junction systems by following the same
procedure as in Sec.~V of \ocite{prb}. Assuming the same gaps and distribution functions in all superconducting
elements, we simply need to substitute
\be
E_J \left|A^d_{s(c)} \right|^2 \to \sum_{j=0}^{M} E_{Jj} \left|A^d_{s(c),j} \right|^2
\ee
in \esref{Gp_sc} and \rref{Gp_sl} (and hence in subsequent equations in Sec.~\ref{sec:nebr}). Here
index $j$ denotes the $M+1$ junctions with Josephson energy $E_{Jj}$ and capacitance $C_j$,
while the matrix elements are defined by
\be
A^d_{s,j} = \frac{1}{2} \left(\langle 1| \sin\frac{\hat\varphi_j}{2} |1 \rangle
- \langle 0| \sin\frac{\hat\varphi_j}{2} |0 \rangle \right)
\ee
with $\varphi_j$ the phase difference across junction $j$. The similar definition for $A^d_{c,j}$ is obtained
by replacing sine with cosine. We remind that the phases are not independent, as they are constrained by
the flux quantization condition
\be
\sum_{j=0}^M \varphi_j = 2\pi f \, , \qquad f =\Phi_e/\Phi_0 \, .
\ee
Below we consider explicitly the two-junction split transmon, while the
many-junction fluxonium is analyzed in Appendix~\ref{app:fx}.

\subsection{Split transmon}
\label{sec:st}

\begin{figure}
\includegraphics[width=0.47\textwidth]{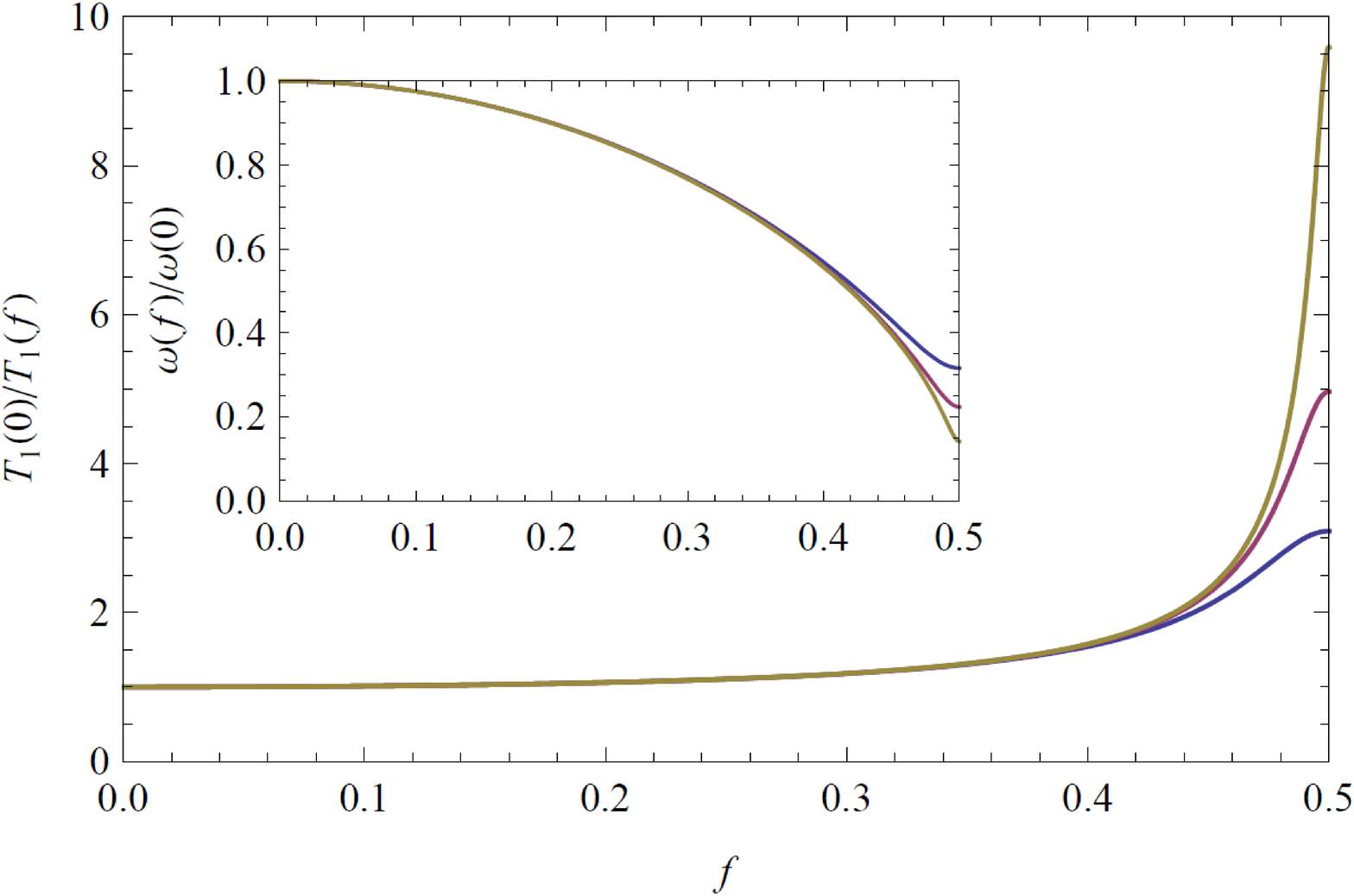}
\caption{Normalized relaxation rate $T_1(0)/T_1(f)$ vs. reduced flux $f$ for (top to bottom) $d=0.02$,
0.05, 0.1. Inset: normalized frequency $\omega_p(f)/\omega_p(0)$ vs. reduced flux for the same values
of the asymmetry parameter (but decreasing top to bottom).}
\label{Fig1}
\end{figure}

The split transmon single degree of freedom is governed by the same Hamiltonian of the single-junction
transmon, but the SQUID loop has a flux-dependent effective Josephson energy
\be\label{ejf}
E_J(f) = \left(E_{J0}+E_{J1}\right) \cos\left(\pi f\right) \sqrt{1+d^2\tan^2\left(\pi f\right)}
\ee
with
\be
d= \frac{\left|E_{J0}-E_{J1}\right|}{E_{J0}+E_{J1}}
\ee
quantifying the junction asymmetry.
In quasi-equilibrium at the effective temperature $T_e$, the relaxation time is given by\cite{prb}
\be\label{st_rel}
\frac{1}{T_1(f)} = \sqrt{\frac{T_e}{\pi \omega_p(f)}} \, e^{-\Delta/T_e}
\frac{\omega_p^2(f)+\omega_p^2(0)}{\omega_p(f)}
\ee
where
\be
\omega_p(f) = \sqrt{8 E_C E_J(f)} \, , \qquad E_C = \frac{e^2}{2(C_0+C_1)}
\ee
We note that the smaller the asymmetry, the larger the tunability of the qubit, since
$\omega_p(0)/\omega_p(1/2) = 1/\sqrt{d}$. However, this flexibility comes at the price of enhancing
the relaxation rate, $T_1(0)/T_1(1/2) = (1+d)/(2d^{3/4})$.
In Fig.~\ref{Fig1} we plot the normalized relaxation rate $T_1(0)/T_1(f)$ as a function of reduced flux $f$ for
three values of the asymmetry parameter. We note that the relaxation rate rises by about a factor 1.5
up to $f\sim 0.4$, but can increase sharply for small asymmetry as $f \to 0.5$.

\begin{figure}
\includegraphics[width=0.47\textwidth]{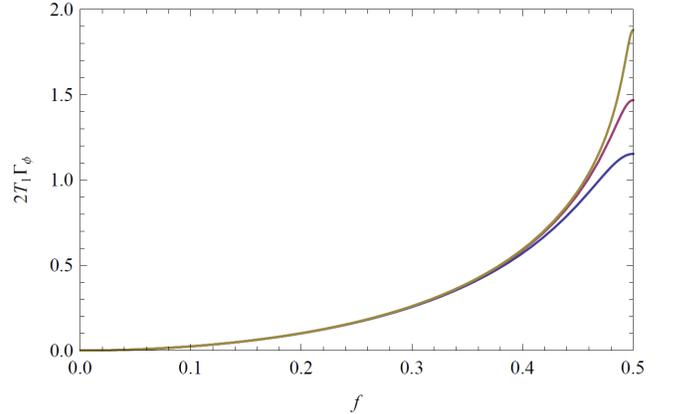}
\caption{Normalized dephasing rate $2T_1 \Gamma_\phi$ vs. reduced flux $f$ for (top to bottom) $d=0.02$,
0.05, 0.1. Other parameters are specified in the text after \eref{st_dep}.
The vanishing of $\Gamma_\phi$ as $f \to 0$ is an artifact of the approximations used to
obtain \eref{st_dep}; a finite dephasing rate at any flux would be obtained by including a subleading
contribution analogous to \eref{tr_dep}.}
\label{Fig2}
\end{figure}

The matrix elements for dephasing are [cf. \eref{ph_me}]
\be\label{st_me}
\left|A^d_{s,j}\right|^2 = \frac{1}{8} \left(\frac{E_C}{\omega_p(f)}\right)^2
\left[1-\cos(\pi f \pm \vartheta)\right]
\ee
where the upper (lower) sign should be used for $j=1$ ($j=0$) and
\be\label{vart}
\tan(\vartheta) = d \tan(\pi f)
\ee
Note that in contrast with the single junction transmon, the matrix elements in general do not vanish
(except at $f=0$). Using \esref{ejf}, \rref{st_me}, and \rref{vart} we find
\be
\sum_{j=0}^1 E_{Jj} \left|A^d_{s,j}\right|^2 = \frac{1}{64} E_C
\left(\frac{\omega_p^2(0)}{\omega_p^2(f)}-1\right)
\ee
and the above-described generalization to multi-junction systems of \eref{Gp_Te} gives
\be\label{st_dep}\begin{split}
\Gamma_\phi = \frac{1}{2\pi} E_C \left(\frac{\omega_p^2(0)}{\omega_p^2(f)}-1\right) e^{-\Delta/T_e}
\\ \times \left[\frac{\Delta}{T_e}+\ln \frac{8\pi T_e}{E_C (\omega_p^2(0)/\omega_p^2(f)-1)} \right]
\end{split}\ee
In Fig.~\ref{Fig2} we show examples of
the dependence of $2T_1 \Gamma_\phi$ on flux for different values of the asymmetry parameter $d$ and
typical values of the other dimensionless parameters ($E_J(0)/E_C = 80$, $\omega_p(0)/\Delta= 0.2$,
$T_e/\Delta =0.06$);
we note that near $f=1/2$ and for small asymmetry, pure dephasing dominates over relaxation,
$2T_1 \Gamma_\phi > 1$. Therefore the pure dephasing effect of quasiparticle
tunneling could be measured in a split transmon if other sources of dephasing (such as flux,
photon, and charge noise) can be suppressed. Charge noise, in particular, can become the dominant
dephasing mechanism as $f\to 1/2$, since the Cooper pair box regime of small $E_J(f)/E_C$ is approached in
this case for small asymmetry.\cite{transmon} However, the contribution of $\Gamma_\phi$ to $1/T_2$
becomes relevant and thus potentially observable at values of reduced flux smaller than $1/2$, where the
system is still in the transmon regime; for example, for $f\sim 0.35$
where $E_J(f)/E_C \sim 0.45 E_J(0)/E_C$, we estimate $2T_1\Gamma_\phi \sim 0.4$.

\section{$T_2^*$ and Andreev states in a Josephson junction}
\label{sec:and}

In the previous sections we have considered the pure dephasing due to the interaction between tunneling
quasiparticles and qubit. Here we study a different quasiparticle mechanism affecting the measured
dephasing rate $1/T_2^*$: as discussed briefly in Sec.~\ref{sec:model} and in more detail in \ocite{prb},
the quasiparticles renormalize the qubit frequency by shifting it by an amount $\delta\omega$
which depends on the quasiparticle occupation. Therefore fluctuations in the
occupation induce frequency fluctuations that can cause additional dephasing.
In this section we focus on the phase qubit and show that this mechanism is not active
during a single measurement, so that it does
not contribute to the pure dephasing rate $\Gamma_\phi$; however, it
can contribute to the time $T_2^*$ by changing the qubit frequency from measurement to
measurement. In other words, this mechanism being slow on the scale of the qubit coherence time,
its dephasing effect can be corrected by using echo techniques.

In a Josephson junction, weakly
bound quasiparticles occupy the Andreev states that carry the dissipationless
supercurrent.\cite{beenakker} Changes in the occupations of these states
affect the value of the critical current (or equivalently of the Josephson energy) and
in turn fluctuations in $E_J$ lead to frequency fluctuations.
As we show below, the parameter determining the relative magnitude
of these fluctuations is the inverse square root of the (effective) number of transmission channels through
the junction; therefore this fluctuation mechanism could be relevant in small junctions.
For each transmission channel $p$ ($p=1,\ldots,N_{ch}$)
with transmission probability $\mathrm{T}_p$ [defined after \eref{HT0}],
we find a corresponding Andreev bound state
with binding energy [see Appendix~\ref{app:abs}]
\be\label{as_be}
\omega^A_p = \Delta - E^A_p \, , \qquad E^A_p = \Delta \left(1- \frac12
\mathrm{T}_p \sin^2\frac{\varphi_0}{2}\right)
\ee
This result is valid for $\mathrm{T}_p \ll 1$; the expression valid for arbitrary
$\mathrm{T}_p$ can be found in \ocite{beenakker}.
The (zero temperature) Josephson energy entering \eref{Hq0} is given by $ E_J = \Delta\sum_p \mathrm{T}_p/4$.
To account for the occupations $x_p^A$ of the Andreev states, due for example to finite temperature,
in \eref{ph_fr} we replace $E_J$ by
\be\label{ej_sub}
E_J \to \frac{\Delta}{4}\sum_{p=1}^{N_{ch}} \mathrm{T}_p \left(1-2x_p^A\right)
\ee
From this substitution we see that a change in the occupation of a single Andreev level
can lead to a small change $\delta E_J$ in the Josephson energy and hence in the qubit frenquency, with a
relative frequency shift of the order of $\delta E_J/E_J \sim 1/N_{ch}$. This effect could be measurable in
small junction ($N_{ch}\sim 10^5$) qubits and may have already been observed in a transmon,
where slow frequency jumps of few parts per million magnitude have been measured.\cite{paik}
More generally we find for the qubit frequency $\omega_q$
at a given set of occupation numbers $x_p^A$
\be\label{om_q}
\omega_q \simeq \EZO  - \frac{8E_C}{\EZO}\cos\varphi_0 \sum_{p=1}^N \frac{\Delta}{4} \mathrm{T}_p x_p^A
\ee
Here we assumed that on average the occupation numbers are small,
$x_\qp^A = \langle x_p^A \rangle \ll 1$;
in quasi-equilibrium the average takes the exponentially small value $x_\qp^A = e^{-\Delta/T_e}$.\cite{xAnote}
From this expression we see that fluctuations of the occupations of the Andreev states lead
to frequency fluctuations.
The mean square fluctuations of $x_p^A$ are related to the average $x_\qp^A$
as\cite{LL5}
\be
\langle \left(\Delta x_p^A\right)^2 \rangle \equiv \langle \left(x_p^A - x_\qp^A\right)^2 \rangle =
 x_\qp^A(1-x_\qp^A)
\ee
Using this expression for the non-degenerate case $x_\qp^A \ll 1$,
we find for the root-mean-square frequency fluctuations
\be\label{rms_fr}
\frac{\sqrt{\langle\left(\Delta \omega_q\right)^2 \rangle}}{\EZO}
 = |\cos\varphi_0| \frac{\omega_p^2}{\EZO^2} \sqrt{x_\qp^A} \frac{1}{\sqrt{N_e}}
\ee
where
\be
N_e = \frac{\left(\sum_p \mathrm{T}_p\right)^2}{\sum_p \mathrm{T}_p^2}
\ee
is the effective number of channels; $N_e$ coincides
with $N_{ch}$ if all the channels have equal transmission probabilities. The number $N_e$
can be estimated independently by measuring the so called subgap structure due to Andreev
reflections,\cite{MARrev}
\be
N_e = \frac{\delta I_1}{\delta I_2}\frac{g_T}{2g_K}\, ,
\ee
where the first factor in the right hand side is the ratio between the current step $\delta I_1$
measured as the voltage increases from below to above $2\Delta/e$ and the subgap current step $\delta I_2$
at $V\sim \Delta/e$.
This ratio is related to junction transparency and is of the order\cite{pekola2,teufel}
$\delta I_2/\delta I_1 \sim 10^{-5}-10^{-3}$, while depending on junction area the ratio
between junction conductance $g_T$ and the conductance quantum $g_K$
is $g_T/g_K \sim 1-100$, so we estimate $N_e \sim 10^3$ to $10^7$
for junction sizes from small to large.

The dephasing effect of the above frequency fluctuations
gives observable contribution to $T_2^*$ if
\be\label{dwqt2}
\langle\left(\Delta \omega_q\right)^2 \rangle^{1/2} T_2 \gtrsim 1 \, .
\ee
Using \eref{rms_fr} this condition is
\be
T_2 \gtrsim \frac{\EZO}{2\omega_p^2} \sqrt{\frac{N_e}{x_\qp^A}} \sim \frac{1}{\omega_p}\sqrt{\frac{N_e}{x_\qp^A}}
\ee
Assuming equilibrium between the occupation factors of Andreev states and free-quasiparticle states
at the effective temperature $T_e \approx 140$~mK (so that $x_\qp^A = e^{-\Delta/T_e}$),
since $\omega_p \sim 10^{11}$~s$^{-1}$ we find $T_2 \gtrsim 10^{-6}$~s ($10^{-4}$~s) for small (large)
juctions. For phase qubits, which are fabricated with large junctions, this estimate is two to
three orders of magnitude longer than the observed coherence time.\cite{Bialczak}
Therefore fluctuations in the occupations of Andreev
levels do not contribute significantly to dephasing in current experiments with phase qubits.

The dephasing effect of the frequency fluctuations can be corrected using an echo pulse if
the occupations do not change during a single measurement. In other words, if the rate at
which the occupations change is small compared to $1/T_2$, then the fluctuations
contribute to the decoherence time $T_2^*$ rather than to $T_2$.
Within our model
Hamiltonian, \eref{Heff}, the only processes that can change the quasiparticle occupations are due to the
interaction between qubit and quasiparticles; for an occupied Andreev level, this interaction leads to its
ionization, with the qubit relaxing and giving its energy to
a bound quasiparticle which is then excited into the continuum part of the spectrum.
Since this process relaxes the qubit, it can in principle contribute to $1/T_1$.
We show in Appendix~\ref{app:abs} that this \textit{intrinsic} contribution is small compared
to the relaxation rate due to the interaction of the qubit with the bulk quasiparticles.
There are of course \textit{extrinsic} mechanisms that could affect the occupations
of the Andreev states and hence the rate of frequency
fluctuations. An example of such a mechanism is flux noise; we estimate that the
ionization rate due to flux noise is in fact small compared to the experimental
$1/T_2$ -- see Appendix~\ref{app:fl_n}.
Another mechanism is the quasiparticle recombination caused by the electron-phonon interaction.
The recombination rate is $\approx x_\qp/\tau_0$, with the characteristic time $\tau_0 \sim
10^{-7}-10^{-6}$~s in aluminum and $\sim 10^{-10}$~s in niobium.\cite{scalapino,wilson}
Since at low temperatures\cite{paik,pekola}
$x_\qp \sim 10^{-7}-10^{-8}$, we find that the recombination rate is much smaller than $1/T_2$.

So far we have considered the effect of fluctuations of the Andreev levels occupations. Other
mechanisms can in principle contribute to decoherence. For example, fluctuations of
the order parameter $\Delta$ in the vicinity of the junction
also affect the Josephson energy, see \eref{ej_sub}; however,
at low temperatures the typical time scale over which $\Delta$ changes in response
to a sudden perturbation is very short, of order $1/\Delta$,\cite{vk}
so these fluctuations do not lead to additional decoherence.
Another mechanism is associated with fluctuations in the number of free (rather than bound)
quasiparticles. As discussed at the end of Sec.~\ref{sec:model}, there are two contributions
to the average frequency shift -- the Josephson one, $\delta\omega_{E_J}$, and the quasiparticle
one, $\delta\omega_\qp$. Fluctuations of free quasiparticle occupations affect the latter, but
their contribution to inhomogeneous broadening is small. Indeed, the average frequency shift can be
obtained by considering the effect of quasiparticles on the junction impedance;\cite{prl,prb}
in quasiequilibrium the contribution
of the normalized quasiparticle density $x_\qp$ to the quasiparticle part $Y_\qp$ of the junction
impedance $Y_J$ is smaller than the term in $Y_J$ proportional to $x_\qp^A$ by the parameter
$\sqrt{T_e/\EZO}$. Moreover, the root mean square fluctuations of $x_\qp$ scale as the inverse
square root of the volume of the electrodes\cite{LL5} and can therefore
be neglected for macroscopic electrodes.

\section{Summary}
\label{sec:summ}

In this work we have
studied decoherence caused by quasiparticles
in superconducting qubits and obtained estimates
for the pure dephasing rate $\Gamma_\phi$ and
for the contribution of inhomogeneous broadening to the
decoherence rate $1/T_2^*$.
We have presented a master equation approach that
not only reproduces and generalizes the formula for the
relaxation rate $1/T_1$ of Refs.~\onlinecite{prl,prb} [see \eref{G10})], but also gives a self-consistent
expression for the pure dephasing rate $\Gamma_\phi$, \eref{Gp_sc}.
Moreover, in studying $1/T_2^*$ we have derived a formula, \eref{rms_fr}, for the typical
fluctuation of the qubit frequency due to change in the occupations of Andreev states.
These two equations are our main results.

Application of \eref{Gp_sc} to single-junction qubits such as the phase qubit, the transmon
(Sec.~\ref{sec:single}), and the flux qubit (Appendix~\ref{app:fl}), and to the
many-junctions fluxonium (Appendix~\ref{app:fx}) shows that in these systems
the pure dephasing rate is a small contribution to decoherence, $2T_1 \Gamma_\phi < 1$. In the split
transmon (Sec.~\ref{sec:st}), on the other hand, the quasiparticle dephasing rate can be larger
than the relaxation rate when the external flux that tunes the qubit frequency
approaches half the flux quantum, see Fig.~\ref{Fig2};
together with its temperature and flux dependence [\eref{st_dep}],
the increased importance of $\Gamma_\phi$ in this regime could permit its experimental measurement.

Finally in Sec.~\ref{sec:and} we have considered the contribution to the decoherence rate $1/T_2^*$ due to
quasiparticles bound into Andreev states localized near the Josephson junction.
Fluctuations of the occupations of these levels from measurement to measurement can in principle induce
dephasing which can be corrected with an echo pulse. In practice, this mechanism gives negligible
contributions to dephasing in current experiments with phase qubits: due to the short observed $T_2$ time,
\eref{dwqt2} implies that the fluctuations of the occupations would need to cause relative frequency
fluctuations of the order $10^{-3}$ to start affecting the coherence of the qubit.

\acknowledgments

This research was funded by Yale University, the Swiss NSF, the Office of the Director
of National Intelligence (ODNI), Intelligence Advanced Research Projects Activity (IARPA),
through the Army Research Office, the American NSF (Contract DMR-1004406), and the DOE
(Contract DE-FG02-08ER46482).

\appendix

\section{Derivation of the master equation}
\label{app:me}

In this Appendix we summarize the main steps of the derivation of the master equation presented in
Sec.~\ref{sec:me}. Our starting point is the von Neumann equation,\cite{petru} which we write
for the two components of the qubit (i.e., reduced) density matrix as
\bea
\label{dzdt}
\frac{d\rho_z}{dt} &=& -i \Tr \left\{ \left[\delta\hat{H}; \hat\rho_{t}\right] \hat\sigma^z \right\} \\
\label{dpdt}
\frac{d\rho_+}{dt} &=& i\EZO \rho_+
 -i \Tr \left\{ \left[\delta\hat{H}; \hat\rho_{t}\right] \hat\sigma^+ \right\}
\eea
Here $\rho_t$ is the total density matrix of the system, comprising both qubit and quasiparticles,
$[\cdot ; \cdot ]$ denotes the commutator and,
as discussed in Sec.~\ref{sec:model}, for our purposes the interaction Hamiltonian
$\delta \hat H= \hat H_T$ is given by \eref{HT} with $A^f_{nm} =0$.
More useful forms of the traces in the right hand sides of the above equations are
\be\label{htzc}\begin{split}
& \Tr \left\{ \left[\hat{H}_T; \hat\rho_{t}\right] \hat\sigma^z \right\} =
\langle\!\langle \left[\hat\sigma^z; \hat{H}_T\right] \rangle\!\rangle
\\ & = 2\tilde{t} \langle\!\langle \left(\hat\sigma^+-\hat\sigma^-\right)
\sum_{n,m,\sigma} A^r_{nm}
\hat\alpha^{L\dagger}_{n\sigma} \hat\alpha^{R}_{m\sigma} \rangle\!\rangle + \mathrm{H.c.}'
\end{split}\ee
and similarly
\be\label{htpmc}\begin{split}
\Tr \left\{ \left[\hat{H}_T; \hat\rho_{t}\right] \hat\sigma^+ \right\} =
\tilde{t} \langle\!\langle \hat\sigma^z
\sum_{n,m,\sigma} A^r_{nm}\hat\alpha^{L\dagger}_{n\sigma} \hat\alpha^{R}_{m\sigma} \rangle\!\rangle
\\ -2\tilde{t} \langle\!\langle \hat\sigma^+
\sum_{n,m,\sigma}A^d_{nm}\hat\alpha^{L\dagger}_{n\sigma} \hat\alpha^{R}_{m\sigma} \rangle\!\rangle
+ \mathrm{H.c.}'
\end{split}\ee
where angular brackets denote quantum statistical averaging with respect to the total density matrix and
the prime denotes that Hermitian conjugation is not applied to qubit operators (i.e.,
Pauli matrices).

The averages in the right hand sides of \esref{htzc}-\rref{htpmc} can be found by solving the equations
governing their time evolution, such as
\be\label{spm_eom}\begin{split}
& -i\partial_t \langle\!\langle \hat\sigma^\pm \hat\alpha^{\dagger L}_{n\sigma} \hat\alpha^R_{m\sigma}
\rangle\!\rangle
= \langle\!\langle \left[\hat{H};\hat\sigma^\pm \hat\alpha^{\dagger L}_{n\sigma} \hat\alpha^R_{m\sigma}\right]
\rangle\!\rangle
\\
& = \left(\pm\EZO + \epsilon^L_n - \epsilon^R_m\right)
\langle\!\langle \hat\sigma^\pm \hat\alpha^{\dagger L}_{n\sigma} \hat\alpha^R_{m\sigma}\rangle\!\rangle
\\ & +
\tilde{t} \bigg\{ \!\pm A^{d*}_{nm}\rho_\pm(t) \left[f^L_n(1-f^R_m) + (1-f^L_n)f^R_m\right]
 -\frac{1}{2} A^{r*}_{nm} \\& \times\left[
\left(1\pm\rho_z(t)\right) f^L_n(1-f^R_m)- (1\mp\rho_z(t)) (1-f^L_n)f^R_m\right]
\bigg\}
\end{split}\ee
The terms in curly brackets originate from averages of one qubit operator times four quasiparticle
operators evaluated in the Born approximation,\cite{petru} for example
\be\begin{split}
& \sum_{i,j,\rho}\langle\!\langle \hat{\sigma}^+ \left\{\hat\alpha^{R\dagger}_{j\rho}\hat\alpha^L_{i\rho};
\hat\alpha^{L\dagger}_{n\sigma} \hat\alpha^R_{m\sigma} \right\} \rangle\!\rangle \\  & = \rho_+(t)
\left[f^L_n(1-f^R_m) + (1-f^L_n)f^R_m\right]
\end{split}\ee
where $\{\cdot ; \cdot \}$ is the anticommutator.
The solution of \eref{spm_eom} is
\be\label{spm_av}\begin{split}
&\langle\!\langle \hat\sigma^\pm \hat\alpha^{\dagger L}_{n\sigma} \hat\alpha^R_{m\sigma}
\rangle\!\rangle = i\tilde{t}\int^t_0\!d\tau \, e^{i(\pm\EZO + \epsilon^L_n - \epsilon^R_m + i 0^+)(t-\tau)}
\\& \bigg\{\!\pm A^{d*}_{nm} \rho_\pm(\tau) \left[f^L_n(1-f^R_m) + (1-f^L_n)f^R_m\right]
-\frac{1}{2} A^{r*}_{nm} \\ & \left[
\left(1\pm\rho_z(\tau)\right) f^L_n(1-f^R_m)- (1\mp\rho_z(\tau)) (1-f^L_n)f^R_m\right]
\bigg\}
\end{split}\ee
A similar expression can be derived for the average in \eref{htpmc} that contains $\hat\sigma^z$.
After substituting these expressions into \esref{htzc}-\rref{htpmc} and the results into
\esref{dzdt}-\rref{dpdt}, we perform two additional approximations. First, we neglect fast rotating terms;
this so-called secular (or rotating wave) approximation\cite{petru} is valid
when the decoherence rate is small on the scale of the qubit frequency, $1/T_2\EZO \ll 1$,
and it amounts to keeping in the
equation for $\rho_z$ only the terms proportional to $(1\pm \rho_z)$ and in the equation for $\rho_+$ only
those proportional to $\rho_+$. With this approximation we find
\be\label{z_eq}\begin{split}
& \frac{d\rho_z(t)}{dt}  =
-2\tilde{t}^2 \int^t_0\!d\tau \sum_{n,m} \left|A^{r}_{nm}\right|^2
\\ & \times \Big\{ \rho_z(\tau) \left[f^L_n(1-f^R_m) + (1-f^L_n)f^R_m\right]
\\ & \quad \times
\Big[ e^{++-}+ e^{-+-} + e^{+-+} + e^{--+}\Big]
\\ &
+
\left[f^L_n(1-f^R_m) - (1-f^L_n)f^R_m\right]
\\ & \quad \times
\Big[ e^{++-}- e^{-+-} - e^{+-+} + e^{--+}\Big]
\Big\}
\end{split}\ee
and
\bea\label{p_eq}
&& \frac{d\rho_+(t)}{dt}  = i\EZO \rho_+(t) \\ \nonumber &&
-2\tilde{t}^2 \int^t_0\!d\tau \sum_{n,m}
\rho_+(\tau) \left[f^L_n(1-f^R_m) + (1-f^L_n)f^R_m\right]
\\ \nonumber && \times \Big\{ 2\left|A^{d}_{nm}\right|^2
\Big[e^{++-} + e^{+-+}\Big]
+ \left|A^{r}_{nm}\right|^2
\Big[e^{0+-} + e^{0-+}\Big]
\Big\}
\eea
where we use the shorthand notation
\be
e^{\alpha\beta\gamma} = e^{i(\alpha\EZO +\beta \epsilon_n^L +\gamma\epsilon_m^R+i0^+)(t-\tau)}
\ee

Next we introduce the Markov approximation\cite{petru} by substituting in the integrands of
\esref{z_eq}-\rref{p_eq}
$\rho_z(\tau) \to \rho_z(t)$, $\rho_+(\tau) \to e^{-i\EZO(t-\tau)}\rho_+(t)$ and extending
the lower integration limits from $0$ to $-\infty$. Then the $\tau$-integrals can be performed using
the identity
\be\label{pd_id}
\int_{-\infty}^t\!\!d\tau \, e^{i(\omega+i0^+) (t-\tau)} = i P \frac{1}{\omega} + \pi \delta(\omega)
\ee
where $P$ denotes the principal part. We note that in \eref{z_eq} the contributions
of the principal parts cancel out, while after rewriting the summations over
$n$, $m$ as integrals over the quasiparticle energies the $\delta$-functions can be used to eliminate one
of these integrals. Assuming equal gaps in the leads, we finally arrive at \eref{rz_me}.

Applying the above steps to \eref{p_eq}, we find that the principal parts cancel out in
the term proportional to $A^d_{nm}$; in that term we assume different gaps
with $\Delta_R> \Delta_L$ to get expression \rref{Gp_def} for the pure dephasing rate
$\Gamma_\phi$. On the other hand, we can take the gaps to be the same in the term
proportional to $A^r_{nm}$; then the $\delta$-functions give rise to the contribution
$-1/2T_1 \rho_+$ in \eref{rp_me}. As for the principal parts, they contribute a term
$
i\delta\tilde{\omega} \rho_+(t)
$
with
\be\label{dto}
\delta\tilde{\omega} = \left|A^r_s\right|^2 \left[F_\qp(-\EZO) - F_\qp(\EZO)\right]
\ee
The function $F_\qp$ is defined in Appendix~A of \ocite{prb}; as in that work, we have
neglected here contributions suppressed by the factor $\EZO/\Delta$. We note that while
$\delta\tilde{\omega}$ has a structure similar to that of $\delta\omega_\qp$ in \ocite{prb}, due
to the projection onto the qubit subspace described in Sec.~\ref{sec:model} the expression in \eref{dto}
accounts for virtual transitions between the qubit states only and neglects those to other
states of the full system. In systems with small anharmonicity (e.g., the transmon and phase qubit)
these transitions cannot be neglected and the average frequency shift must be calculated using the formulas
in \ocite{prb}. Finally, we remind that the total average frequency shift $\delta\omega$ contains also a
Josephson part $\delta \omega_{E_J}$, as discussed in Sec.~\ref{sec:model}.

\section{Dephasing at next-to-leading order}
\label{app:scd}

The self-consistent equation \rref{Gp_sc} for $\Gamma_\phi$ requires going beyond the lowest order
(in the tunneling amplitude $\tilde{t}$) perturbative considerations of Appendix~\ref{app:me} in order
to regularize the logarithmic divergence in \eref{Gp_def} for equal gaps.
Here we focus on the next to leading order contributions to validate that equation. First, however,
let us discuss briefly the smearing of the singularity, \eref{smearing}, which is obtained as follows:
after the Markov approximation, the term in \eref{p_eq} proportional to $A^d_{nm}$ is explicitly
\be\begin{split}
-4\tilde{t}^2 \rho_+(t)  \sum_{n,m}
\left[f^L_n(1-f^R_m) + (1-f^L_n)f^R_m\right]
 \left|A^{d}_{nm}\right|^2 \\ \lim_{\gamma \to 0^+} \int^t_{-\infty}\!d\tau
\left[e^{i(\epsilon^L_n-\epsilon_m^R+i\gamma)(t-\tau)} + e^{i(-\epsilon^L_n+\epsilon_m^R+i\gamma)(t-\tau)}\right]
\end{split}\ee
Rather than taking the limit, we assume $\gamma$ small but finite
(in particular, $\gamma \ll \EZO$ for the rotating wave approximation to be valid). After
integration the last line becomes
\be
\frac{2\gamma}{(\epsilon^L_n - \epsilon^R_m)^2 + \gamma^2}
\ee
This explains the origin of the last factor in the second line of \eref{smearing}, with the other
factors accounting for the square root singularity of the BCS density of states. We now want to show that the
identification $\gamma = \Gamma_\phi$ is correct at next to leading order. To do so, we initially assume that
the left/right gaps are different, so that the logarithmic divergence is absent and the perturbative expansion
in $\tilde{t}$ is justified. Next, we keep only those terms that would become
logarithmically divergent in the limit of equal gaps.

To begin our derivation, we note that
in \eref{p_eq} the first term in square brackets multiplying $A^d_{nm}$ originates from
$\langle\!\langle \hat\sigma^+ \hat\alpha^{\dagger L}_{n\sigma} \hat\alpha^R_{m\sigma}
\rangle\!\rangle$, as explained in Appendix~\ref{app:me}. Together with
the other term in square brackets, they give rise to the pure dephasing rate term in the master
equation \rref{p_eq} via the equality
\bea\label{avgp}
2\tilde{t}\sum_{n,m,\sigma}\Big[
A^d_{nm} \langle\!\langle \hat\sigma^+ \hat\alpha^{L\dagger}_{n\sigma} \hat\alpha^R_{m\sigma}\rangle\!\rangle
&+&
A^{d*}_{nm} \langle\!\langle \hat\sigma^+ \hat\alpha^{R\dagger}_{m\sigma} \hat\alpha^L_{n\sigma} \rangle\!\rangle
\Big] \nonumber \\ && \qquad = i\Gamma_\phi \rho_+(t)
\eea
In what follow we first consider in some detail the next order contributions to
$\langle\!\langle \hat\sigma^+ \hat\alpha^{\dagger L}_{n\sigma} \hat\alpha^R_{m\sigma}
\rangle\!\rangle$ and then discuss briefly the contributions to other averages.
Without invoking the lowest order Born approximation, the equation of motion for
$\langle\!\langle \hat\sigma^+ \hat\alpha^{\dagger L}_{n\sigma} \hat\alpha^R_{m\sigma}
\rangle\!\rangle$ is obtained by adding to the right hand side of \eref{spm_eom} the terms
\be\label{ntl}\begin{split}
\tilde{t} \sum_{i,j,\rho} \bigg[A^d_{ij}N^{\sigma,\rho}_{nm,ij} + A^{d*}_{ij} M^{\sigma,\rho}_{nm,ij}
-\frac12 A^r_{ij}Q^{\sigma,\rho}_{nm,ij} \\ -\frac12 A^{r*}_{ij} P^{\sigma,\rho}_{nm,ij}
+\frac12 A^{r*}_{ij} S^{\sigma,\rho}_{nm,ij}+\frac12 A^{r*}_{ij} R^{\sigma,\rho}_{nm,ij}\bigg]
\end{split}\ee
with the definitions
\bea\label{m_def}
M^{\sigma,\rho}_{nm,ij} &=& \langle\!\langle \hat\sigma^+ \left\{\hat\alpha^{R\dagger}_{j\rho}
\hat\alpha^L_{i\rho}; \hat\alpha^{L\dagger}_{n\sigma} \hat\alpha^R_{m\sigma} \right\} \rangle\!\rangle \\
 \nonumber &-&
\delta_{ni} \delta_{mj} \delta_{\sigma\rho} \rho_+ \left[f^L_n(1-f^R_m) + (1-f^L_n)f^R_m\right] \\
\label{n_def}
N^{\sigma,\rho}_{nm,ij} &=& \langle\!\langle \hat\sigma^+ \left\{\hat\alpha^{L\dagger}_{i\rho}
\hat\alpha^R_{j\rho}; \hat\alpha^{L\dagger}_{n\sigma} \hat\alpha^R_{m\sigma} \right\} \rangle\!\rangle
\eea
\bea\label{p_def}
P^{\sigma,\rho}_{nm,ij} &=& \langle\!\langle \hat\sigma^z \left\{\hat\alpha^{R\dagger}_{j\rho}
\hat\alpha^L_{i\rho}; \hat\alpha^{L\dagger}_{n\sigma} \hat\alpha^R_{m\sigma} \right\} \rangle\!\rangle \\
\nonumber &-& \delta_{ni} \delta_{mj} \delta_{\sigma\rho} \rho_z \left[f^L_n(1-f^R_m) + (1-f^L_n)f^R_m\right]\\
Q^{\sigma,\rho}_{nm,ij} &=& \langle\!\langle \hat\sigma^z \left\{\hat\alpha^{L\dagger}_{i\rho}
\hat\alpha^R_{j\rho}; \hat\alpha^{L\dagger}_{n\sigma} \hat\alpha^R_{m\sigma} \right\} \rangle\!\rangle
\eea
\bea
R^{\sigma,\rho}_{nm,ij} &=& \langle\!\langle \left[\hat\alpha^{R\dagger}_{j\rho}\hat\alpha^L_{i\rho};
\hat\alpha^{L\dagger}_{n\sigma} \hat\alpha^R_{m\sigma} \right] \rangle\!\rangle\\ \nonumber &-&
\delta_{ni} \delta_{mj} \delta_{\sigma\rho} \left[f^R_m -f^L_n \right]\\
\label{s_def}
S^{\sigma,\rho}_{nm,ij} &=& \langle\!\langle \left[\hat\alpha^{L\dagger}_{i\rho}\hat\alpha^R_{j\rho};
\hat\alpha^{L\dagger}_{n\sigma} \hat\alpha^R_{m\sigma} \right] \rangle\!\rangle
\eea
In introducing these definitions we have subtracted out the lowest order contributions already appearing
in \eref{spm_eom}. Then in that equation and in \esref{m_def} and \rref{p_def} the density matrix should
be understood as the lowest (zeroth) order one. In other words,
by construction the quantities defined in \esref{m_def}-\rref{s_def} account for higher order (in $\tilde{t}$)
contributions; these can be found by considering the equations of motions for those quantities, such as
\be\label{meq0}\begin{split}
-i\partial_t M_{nm,ij}^{\sigma,\rho} = (\EZO + \epsilon^L_n - \epsilon^R_m +\epsilon^R_j - \epsilon^L_i)
M_{nm,ij}^{\sigma,\rho} \\
+\tilde{t} \sum_{k,l,\mu} \langle\!\langle \hat\sigma^+ \left\{A^d_{kl} \hat\alpha^{L\dagger}_{k\mu}
\hat\alpha^R_{l\mu} + A^{d*}_{kl} \hat\alpha^{R\dagger}_{l\mu} \hat\alpha^L_{k\mu};{\cal A}_{nm,ij}^{\sigma,\rho}
\right\} \\ -\frac12 \hat\sigma^z \left\{A^r_{kl} \hat\alpha^{L\dagger}_{k\mu}
\hat\alpha^R_{l\mu} + A^{r*}_{kl} \hat\alpha^{R\dagger}_{l\mu} \hat\alpha^L_{k\mu};{\cal A}_{nm,ij}^{\sigma,\rho}
\right\} \\ +\frac12 \left[A^r_{kl} \hat\alpha^{L\dagger}_{k\mu}
\hat\alpha^R_{l\mu} + A^{r*}_{kl} \hat\alpha^{R\dagger}_{l\mu} \hat\alpha^L_{k\mu};{\cal A}_{nm,ij}^{\sigma,\rho}
\right]
\rangle\!\rangle
\end{split}\ee
where ${\cal A}_{nm,ij}^{\sigma,\rho}$ stands for the anticommutator
\be
{\cal A}_{nm,ij}^{\sigma,\rho} = \left\{\hat\alpha^{R\dagger}_{j\rho}\hat\alpha^L_{i\rho};
\hat\alpha^{L\dagger}_{n\sigma} \hat\alpha^R_{m\sigma} \right\}
\ee
At lowest order, all the averages in the right hand side of \eref{meq0} vanish; non-vanishing
contributions can in principle be found by considering once again the equation of motions for those
averages. As it is well known, proceeding in this manner we would obtain a hierarchy of coupled
equations.\cite{ms} Here we make two approximations: first, we truncate the hierarchy at this level;
second, as explained above we keep only those terms that in the limit of equal gaps
would give logarithmically divergent contributions to the master equation. As a first step, this amounts
to performing a mean-field like approximation
in which the averages in the right hand side of \eref{meq0} are written in terms of product of averages
as in the following example:
\be\begin{split}
\langle\!\langle \hat\sigma^+ \left\{\hat\alpha^{L\dagger}_{k\mu} \hat\alpha^R_{l\mu} ;
{\cal A}_{nm,ij}^{\sigma,\rho} \right\} \rangle\!\rangle & =
2 \langle\!\langle \hat\sigma^+ \hat\alpha^{L\dagger}_{k\mu} \hat\alpha^R_{l\mu}  \rangle\!\rangle
\langle\!\langle {\cal A}_{nm,ij}^{\sigma,\rho} \rangle\!\rangle \\
& + 2 \langle\!\langle \hat\sigma^+ \hat\alpha^{L\dagger}_{n\sigma} \hat\alpha^R_{m\sigma} \rangle\!\rangle
\langle\!\langle  {\cal A}_{kl,ij}^{\mu,\rho} \rangle\!\rangle
\end{split}\ee
where
\be
\langle\!\langle {\cal A}_{nm,ij}^{\sigma,\rho} \rangle\!\rangle  =
\delta_{ni}\delta_{mj} \delta_{\sigma\rho}
\left[f^L_n(1-f^R_m) + (1-f^L_n)f^R_m\right]
\ee
Similar expressions can be written for the other averages appearing in \eref{meq0}.
In the second step we check which of the terms obtained in this way are logarithmically divergent in the limit
of equal gaps and discard those that are finite (here we employ again the Born-Markov\cite{hma} and
rotating wave approximations).

Applying the above procedure to \eref{meq0} we find that the terms
in the last two lines can be neglected, while in terms originating
from the second line we use \eref{avgp} as well as \eref{spm_av} (in
the rotating wave approximation, we only need to keep the term in
the right hand side of that equation  that contains $\rho_+$). Solving
the equation for $M_{nm,ij}^{\sigma,\rho}$ so obtained we finally
arrive at
\begin{widetext}
\be\label{mres}\begin{split}
& M_{nm,ij}^{\sigma,\rho} (t) = - t \Gamma_\phi \rho_+(t) \
\delta_{ni} \delta_{mj} \delta_{\sigma\rho} \left[f^L_n(1-f^R_m) + (1-f^L_n)f^R_m\right]
-2\tilde{t}^2 \rho_+(t) A^d_{ij} A^{d*}_{nm} \left[f^L_i(1-f^R_j) + (1-f^L_i)f^R_j\right]
\\ & \times
\left[f^L_n(1-f^R_m) + (1-f^L_n)f^R_m\right]
\int^t_0\!d u \
e^{i(\epsilon^L_n - \epsilon^R_m +\epsilon^R_j - \epsilon^L_i + i0^+)(t-u)}
\int^u_0\!d\tau \left( e^{i(\epsilon^L_n - \epsilon^R_m + i 0^+)(u-\tau)} +
 e^{i(-\epsilon^L_i + \epsilon^R_j + i 0^+)(u-\tau)}\right)
\end{split}\ee
\end{widetext}
We then use the same approach to find the expression for $N_{nm,ij}^{\sigma,\rho}$ [\eref{n_def}], which
has the structure similar to that of the last term in \eref{mres}. Using these results we get
\be\label{mn_sum}\begin{split}
& \sum_{i,j,\rho} \left(A^d_{ij} N_{nm,ij}^{\sigma,\rho}
+ A^{d*}_{ij} M_{nm,ij}^{\sigma,\rho}\right) \\
& = - \Gamma_\phi \rho_+(t)
 A^{d*}_{nm} \left[f^L_n(1-f^R_m) + (1-f^L_n)f^R_m\right] \\ & \qquad \times
\left\{t + \int^t_0\!d\tau \ e^{i(\epsilon^L_n - \epsilon^R_m + i0^+)(t-\tau)} \right\}
\end{split}\ee
To obtain the last term in curly brackets we used the identity
\be\begin{split}
\int^t_0\!d u \int^u_0\!d\tau  \ h(\tau, u)
& = \int^t_0\!d\tau \int^t_0 \!d u \ h(\tau, u) \\ & -
 \int^t_0\!d u \int^u_0 \!d\tau \ h(u,\tau)
\end{split}\ee
to combine contributions coming from $M_{nm,ij}^{\sigma,\rho}$ and $N_{nm,ij}^{\sigma,\rho}$ in a compact form.

Using the same procedure one can find the expressions for the quantities defined in \esref{p_def}-\rref{s_def}.
Those quantities, however, do not contribute to the master equation within the approximations we are
employing (in particular, we remind that in the rotating wave approximation we neglect by assumption terms
small by the factor $\Gamma_\phi/\EZO$). Therefore, we obtain the following
next-to-leading order equation of motion for
$\langle\!\langle \hat\sigma^+ \hat\alpha^{\dagger L}_{n\sigma} \hat\alpha^R_{m\sigma}
\rangle\!\rangle$ by substituting \eref{mn_sum} into \eref{ntl} and adding the result to the left
hand side of \eref{spm_eom}:
\be\label{spm_eom2}\begin{split}
& -i\partial_t \langle\!\langle \hat\sigma^+ \hat\alpha^{\dagger L}_{n\sigma} \hat\alpha^R_{m\sigma}
\rangle\!\rangle
 = \left(\EZO + \epsilon^L_n - \epsilon^R_m \right)
\langle\!\langle \hat\sigma^+ \hat\alpha^{\dagger L}_{n\sigma} \hat\alpha^R_{m\sigma}\rangle\!\rangle
\\ & +
\tilde{t} A^{d*}_{nm}\rho_+\left(1-\Gamma_\phi t -
\Gamma_\phi \int^t_0\!d\tau \ e^{i(\epsilon^L_n - \epsilon^R_m + i0^+)(t-\tau)}\right)
\\ & \qquad \times \left[f^L_n(1-f^R_m) + (1-f^L_n)f^R_m\right]
\\ & -\frac{\tilde{t}}{2} A^{r*}_{nm}
\left[ \left(1+\rho_z\right) f^L_n(1-f^R_m)- (1-\rho_z) (1-f^L_n)f^R_m\right]
\end{split}\ee
As explained at the beginning of this Appendix, we want to show that this equation
agrees at next to leading order with the smearing obtained by introducing a finite decay rate in the
terms responsible for dephasing, with the decay rate given by $\Gamma_\phi$ itself. Indeed,
introducing this decay in \eref{spm_av} we find
\be\begin{split}
&\langle\!\langle \hat\sigma^+ \hat\alpha^{\dagger L}_{n\sigma} \hat\alpha^R_{m\sigma}
\rangle\!\rangle = i\tilde{t} A^{d*}_{nm}
\int^t_0\!d\tau \, e^{i(\EZO + \epsilon^L_n - \epsilon^R_m + i \Gamma_\phi)(t-\tau)}
\\&  \rho_+(\tau) \left[f^L_n(1-f^R_m) + (1-f^L_n)f^R_m\right]
\\ & -\frac{1}{2} i \tilde{t} A^{r*}_{nm}
\int^t_0\!d\tau \, e^{i(\EZO + \epsilon^L_n - \epsilon^R_m + i 0^+)(t-\tau)} \\ &
\left[\left(1+\rho_z(\tau)\right) f^L_n(1-f^R_m)- (1-\rho_z(\tau)) (1-f^L_n)f^R_m\right]
\end{split}\ee
Taking the time derivative of this equation we get
\be\label{sp_eoma}\begin{split}
& -i\partial_t \langle\!\langle \hat\sigma^+ \hat\alpha^{\dagger L}_{n\sigma} \hat\alpha^R_{m\sigma}
\rangle\!\rangle
 = \left(\EZO + \epsilon^L_n - \epsilon^R_m \right)
\langle\!\langle \hat\sigma^+ \hat\alpha^{\dagger L}_{n\sigma} \hat\alpha^R_{m\sigma}\rangle\!\rangle
\\ & +
\tilde{t} A^{d*}_{nm}\rho_+ \left[f^L_n(1-f^R_m) + (1-f^L_n)f^R_m\right]
\\ & -\frac{\tilde{t}}{2} A^{r*}_{nm}
\left[ \left(1+\rho_z\right) f^L_n(1-f^R_m)- (1-\rho_z) (1-f^L_n)f^R_m\right] \\
& -\Gamma_\phi \tilde{t}  A^{d*}_{nm}
\int^t_0\!d\tau \, \rho_+(\tau) e^{i(\EZO + \epsilon^L_n - \epsilon^R_m + i \Gamma_\phi)(t-\tau)}
\\& \qquad \times  \left[f^L_n(1-f^R_m) + (1-f^L_n)f^R_m\right]
\end{split}
\ee
At next-to-leading order, one should expand the exponentially decaying part of $\rho_+$
[cf. \esref{p_t}-\rref{t2def}] in the second line of \eref{sp_eoma} and hence substitute there, with
logarithmic accuracy, $\rho_+ \to \rho_+ (1-\Gamma_\phi t)$. The last term in \eref{sp_eoma}
is explicitly of higher order, so one can use $\rho_+(\tau) \simeq e^{i\EZO \tau}$ and drop
$\Gamma_\phi$ in the exponent. In this way we recover \eref{spm_eom2}, thus showing for
$\langle\!\langle \hat\sigma^+ \hat\alpha^{\dagger L}_{n\sigma} \hat\alpha^R_{m\sigma}
\rangle\!\rangle$ the validity of our Ansatz.
To complete the proof, we repeat the above steps for other averages, such as
$\langle\!\langle \hat\sigma^+ \hat\alpha^{\dagger R}_{m\sigma} \hat\alpha^L_{n\sigma}
\rangle\!\rangle$ and $\langle\!\langle \hat\sigma^z \hat\alpha^{\dagger L}_{n\sigma} \hat\alpha^R_{m\sigma}
\rangle\!\rangle$. The latter contributes to the $1/2T_1$ term in the master equation
\rref{rp_me} and at next-to-leading order the only correction we find is that corresponding to the expansion of
the exponentially decaying part of $\rho_+$, as discussed above for the second line in \eref{sp_eoma}.

\section{Phase relaxation in flux qubit and fluxonium}
\label{app:flfl}

\subsection{Flux qubit}
\label{app:fl}

In a flux qubit, the external flux threading the superconducting loop is tuned to half the flux quantum,
$f = \Phi_e/\Phi_0 \simeq 1/2$, and the potential energy takes the form of a double well.
Then the qubit states $|\pm \rangle$ are the two lowest tunnel-split states in this potential
with energy difference
\be\label{fl_fr}
\EZO(f) = \sqrt{\bar{\epsilon}^2+ \left[(2\pi)^2 \bar{E}_L(f-1/2)\right]^2}
\ee
where for $\bar{E}_J \gg \bar{E}_C$ we have
\be\label{eps_def}
\bar{\epsilon} = 2\sqrt{\frac{2}{\pi}} \sqrt{8\bar{E}_C \bar{E}_J}\left(\frac{8\bar{E}_J}{\bar{E}_C}\right)^{1/4}
e^{-\sqrt{8\bar{E}_J/\bar{E}_C}}
\ee
Expressions for the renormalized parameters $\bar{E}_C$, $\bar{E}_J$, and $\bar{E}_L$ in terms of the bare
parameters of the Hamiltonian \rref{Hq0} can be found in Sec.~IV.B of \ocite{prb}. It was shown there that
the matrix element $A^r_s$ vanishes at $f=1/2$ because of symmetry considerations, thus leading to a
minimum for the relaxation rate. Here we focus on the case $f=1/2$ and therefore we
need to evaluate the contribution to relaxation originating from the last line in \eref{G10}. The relevant
matrix element is
\be
\left|A^r_c\right| = \frac{\bar{\epsilon}}{\EZO(f)}
\ee
which equals unity at $f=1/2$. Then from \eref{G10} we obtain
\be\label{fl_rel}
\frac{1}{T_1} = \frac{8}{\pi} E_J \sqrt{\frac{\EZO}{2\Delta}} x_\qp = \frac{8}{\pi} E_J
\sqrt{\frac{\pi \bar{\epsilon} T_e}{\Delta^2}} \, e^{-\Delta/T_e}
\ee

Turning now to the dephasing rate, we find at $f=1/2$ the following expression for the matrix element
\be\label{fl_me}
\left|A^d_s\right| = \frac{D}{2\sqrt{2}} \frac{\bar{\epsilon}}{\bar{E}_J}
\left(\frac{\bar{E}_J}{\bar{E}_C}\right)^{1/3},
\ee
where $D\approx 1.45$ is a numerical coefficient.\cite{prb} Using \eref{Gp_Te} and \rref{fl_rel}, after
straightforward algebra we arrive at
\be\begin{split}
2T_1 \Gamma_\phi = \frac{D^2}{\sqrt{\pi}}\sqrt{\frac{\Delta}{E_C}}\sqrt{\frac{\Delta}{T_e}}
\left(\frac{\bar{\epsilon}}{\bar{E}_C}\right)^{3/2} \left(\frac{\bar{E}_C}{\bar{E}_J}\right)^{4/3}
\\ \times \left\{\frac{\Delta}{T_e}+ \ln \left[\frac{\pi}{D^2}\frac{T_e}{\Delta}\frac{\Delta}{\bar{E}_C}
\left(\frac{\bar{E}_C}{\bar{\epsilon}}\right)^2
\left(\frac{\bar{E}_J}{\bar{E}_C}\right)^{1/3}\right]  \right\}
\end{split}\ee
Due to the exponential suppression of the splitting, \eref{eps_def}, this quantity is in general small.
Indeed, for $\bar{E}_C/\Delta, \, T_e/\Delta > 0.01$ and $\bar{E}_J/\bar{E}_C \gtrsim 15$ we find
$2T_1 \Gamma_\phi \lesssim 0.01$.

\subsection{Fluxonium}
\label{app:fx}

In the fluxonium an array of $M\gg 1$ identical junctions
(each with Josephson energy $E_{J1} \gg E_{C1}$ large compared to their charging energy)
acts as a lossy inductor connected to a weaker junction with $E_{J0} < E_{J1}$. The inductive energy
of the array is $E_L = E_{J1}/M$ and the losses are due to quasiparticle tunneling through the
array junctions. In fact, for external flux near half the flux quantum the relaxation time is determined
by this loss mechanism,\cite{prb}
\be\label{fx_rel}
\frac{1}{T_1} = 4\pi E_L \sqrt{\frac{\pi T_e}{\EZO(f)}} \, e^{-\Delta/T_e}
\left(\frac{\EZO(1/2)}{\EZO(f)}\right)^2 ,
\ee
since as discussed above for the flux qubit the contribution of the weaker junction is
suppressed at $f=1/2$ [cf. \eref{fl_rel}]. Note that at $f=1/2$ the rate in \eref{fx_rel} is
larger than that in \eref{fl_rel} by the factor $(\Delta/E_J)(E_L/\EZO(1/2))$.

To calculate the dephasing rate, we note that at $f=1/2$ the matrix element for the weak junction is the
same as for the flux qubit, \eref{fl_me},
\be
\left|A^d_{s,0} \right| = \frac{D}{2\sqrt{2}} \frac{\EZO(1/2)}{\bar{E}_{J0}}
\left(\frac{\bar{E}_{J0}}{\bar{E}_{C0}}\right)^{1/3}
\ee
while for each array junction we get
\be
\left|A^d_{s,1} \right| = \frac{\pi}{2M} \left|A^d_{s,0} \right|
\ee
Then the coefficient containing the sum over all junctions is
\be
\sum_{j=0}^{M} E_{Jj} \left|A^d_{s,j} \right|^2 =
\left(E_{J0} + \frac{\pi^2}{4} E_L \right) \left|A^d_{s,0} \right|^2,
\ee
which in the limit $\bar{E}_J/\bar{E}_C \gg 1$ is exponentially suppressed, see \eref{eps_def}.
Therefore at $f=1/2$ the dephasing rate $\Gamma_\phi$
has the same exponential suppression as in the flux qubit. Since as discussed above the fluxonium relaxation
rate is parametrically larger than the flux qubit one,\cite{fl_perf} we find again that
the pure dephasing rate is small compared to the relaxation rate for large $\bar{E}_J/\bar{E}_C $.
The latter condition is not satisfied experimentally, since
typically\cite{fx_exp} $E_J/E_C \lesssim 5$, and numerical calculations beyond the scope
of the present work may be needed to address this parameter regime. However, we note
that in all cases studied here decreasing the ratio $E_J/E_C$ increases the relative contribution
of pure dephasing to $1/T_2$.

\section{Andreev bound states and ionization rate}
\label{app:abs}

The goals of this Appendix are to derive \esref{as_be} starting from the model defined
by \esref{Heff}-\rref{HT0}, and to estimate the ionization rates due
to qubit-quasiparticles interaction and flux noise. In the low energy limit where the
characteristic energy of
the quasi-particles $\delta E$ as well as the qubit transition
frequency $\EZO$ are small compared to the superconducting gap
$\Delta$, we approximate the
BCS coherence factors as $u_{n}^j\approx v_n^j\approx 1/\sqrt{2}$.
Then considering for now a single channel junction, \eref{HT0} takes the form\cite{prb}
\begin{equation}
\hat H_T=i\tilde
t\sin(\hat\varphi/2)\sum_{n,m,\sigma}\hat\alpha_{n\sigma}^{L\dagger}\hat\alpha_{m\sigma}^R+{\rm
  H.c.}
\end{equation}
Assuming for simplicity identical left/right leads, we perform a canonical rotation into a new
quasiparticle basis defined by the operators
\be
\hat\gamma_{\pm n\sigma}=\frac{1}{\sqrt{2}}\left(\hat\alpha_{n\sigma}^L\pm i\hat\alpha^R_{n\sigma}\right).
\ee
In this basis we have [cf. \eref{Hqp}]
\begin{align}
\hat H_\qp&= \hat H_{\qp +} + \hat H_{\qp -}\, , \quad \hat H_{\qp \pm} =
\sum_{n,\sigma}\epsilon_n \hat\gamma_{\pm n\sigma}^{\dagger}\hat\gamma_{\pm n\sigma} \label{hqppm} \\
\hat H_T &=\tilde t\sin\frac{\hat\varphi}{2}
\sum_{n,m,\sigma}\left(\hat\gamma_{-n\sigma}^{\dagger}\hat\gamma_{-m\sigma}
-\hat\gamma_{+n\sigma}^{\dagger}\hat\gamma_{+m\sigma}\right).
\end{align}
Denoting with $\ket{j}$ and $\mathcal{E}_j$ the eigenstates and eigenenergies of $\hat H_{\varphi}$ [\eref{Hq0}],
the total Hamiltonian $\hat H$ can then be split into
parts that are respectively diagonal and non-diagonal in the qubit
subspace, $\hat H=\hat H_{\rm d} +\hat H_{\rm nd}$, with the
diagonal part defined as
\be
\hat H_{\rm d} =\sum_{j}\mathcal{E}_j\ket{j}\bra{j} +\sum_j\ket{j}\bra{j}\left(\hat
  H_{j+}+\hat H_{j-}\right),
\ee
where
\be\label{Hjpm}
\hat H_{j\pm}=
\hat H_{\qp \pm}
\mp \tilde{t} s_{jj}\sum_{n,m,\sigma}\hat\gamma_{\pm n\sigma}^{\dagger}\hat\gamma_{\pm m\sigma}.
\ee
and we have used the definition \rref{medef} for the matrix elements $s_{ij}$.
The non-diagonal part is given by
\begin{align}\label{eq:2}
\hat H_{\rm nd}&=\tilde t\,\sum_{i\neq j} s_{ij}\ket{i}\bra{j}\sum_{m,n,\sigma}
\left(\hat\gamma_{-n\sigma}^{\dagger}\hat\gamma_{-m\sigma}-
\hat\gamma_{+n\sigma}^{\dagger}\hat\gamma_{+m\sigma}\right)
\end{align}
It describes real transitions in which qubit and quasiparticles exchange energy.
The term proportional to $\tilde{t}$ in the diagonal part, on the other hand, accounts for virtual
transitions that renormalize the spectrum. Indeed,
as we show next, for $s_{jj}>0$ ($s_{jj}<0$) there exists a sub-gap
Andreev bound state in the $\gamma_+$ ($\gamma_-$)
subspace. Because the two subspaces are uncoupled,
we can restrict ourselves to either one of those; in the following we
consider the $\gamma_+$ subspace.

Since $\hat{H}_\mathrm{d}$ is diagonal in the qubit space, to find the spectrum we only need to
calculate the eigenvalues of the quasiparticle Hamiltonians $\hat{H}_{j\pm}$.
We denote with $\ket{A_j}$ the wavefunction of the Andreev state when the qubit is in state $\ket{j}$;
to solve the Schr\"odinger equation $\hat H_{j+}\ket{A_j}=E\ket{A_j}$ we make the Ansatz
$\ket{A_j}=\sum_{n\sigma}a_{jn}\gamma_{+n\sigma}^{\dagger}\ket{\emptyset}$, where $\ket{\emptyset}$
denotes the quasiparticle vacuum state, $\gamma_{\pm n \sigma} \ket{\emptyset} = 0$, and obtain
the following system of linear equations
\be\label{eq:1}
  a_{jn}=\tilde ts_{jj}\frac{1}{\epsilon_n-E}\sum_m a_{jm}.
\ee
To find the eigenenergy $E$, we sum both sides over $n$ and in the low
energy limit we write $\epsilon_n\approx \Delta + {\xi_n}^2/(2\Delta)$; then the sum over $n$ in the right
hand side can be approximated by an integral, $\sum_n \approx \nu_0 \int d\xi$, and we arrive at
\begin{align}
1 = \pi\nu_0\tilde ts_{jj}\sqrt{\frac{2\Delta}{\Delta - E}}.
\end{align}
A solution with energy $E<\Delta$ exists if and only if
$s_{jj}>0$ (the opposite holds in the $\gamma_-$ subspace where a bound
state exists if and only if $s_{jj}<0$.). The corresponding bound
state energy is
\begin{equation}\label{eaj}
E^A_j
=\Delta[1-2(\pi\nu_0\tilde t)^2{s_{jj}}^2],
\end{equation}
This energy depends on the state of the qubit via the matrix element $s_{jj}$. However,
for the low-energy states of the phase qubit
this matrix element is the same at leading order in $E_C/\EZO \ll 1$, since
the square of the matrix element is\cite{prb}
\be\label{sij_ph}\begin{split}
{s_{ij}}^2 = & \, \delta_{i,j}\left[1 - 2\frac{E_C}{\EZO} \left(i + \frac12\right)\right]
\sin^2\frac{\varphi_0}{2}
\\ &+\frac{E_C}{\EZO}\left[j\delta_{i,j-1} + (j+1) \delta{i,j+1} \right]  \cos^2\frac{\varphi_0}{2}
\end{split}\ee
up to higher order terms $\propto(E_C/\EZO)^2$.
Keeping only the leading term in this equation, introducing the transmission probability
$\mathrm{T} = (2\pi\nu_0 \tilde{t})^2$ in \eref{eaj}, and generalizing it to multiple channels,
we arrive at \eref{as_be}. (In that equation the subscript $p$ denotes the transmission channel,
and we have dropped the qubit state index $j$ since, as explained above, the leading order expression
is independent of $j$.)

For later use, we note that the normalization condition $\sum_n(a_{jn})^2=1/2$, which accounts for spin
degeneracy, together with the square of Eq.~(\ref{eq:1}), leads to the amplitudes
\begin{equation}\label{eq:5}
a_{jn} = \frac{1}{\sqrt{\pi\nu_0}}\frac{(2\Delta\omega^A_j)^{3/4}}{{\xi_n}^2+2\Delta\omega^A_j},
\end{equation}
where $\omega^A_j=\Delta-E^A_j$ is the binding energy.

\subsection{Ionization rate}

The ionization of the Andreev level can be caused
by quantum fluctuations of the phase difference across the junction induced by the finite charging energy
$E_C$; the ionization rate can be calculated using Fermi's golden rule by treating
the non-diagonal part~(\ref{eq:2}) of the Hamiltonian as a perturbation.
For a qubit initially in
the state $\ket{i}$, the ionization rate $\Gamma^A_i$ is given by
\begin{align}\label{eq:3}
\Gamma^A_i&=2\pi\sum_{nj}\left|\bra{j,\epsilon_{jn}} \hat H_{\rm
    nd}\ket{i,A_i}\right|^2\nonumber\\
&\qquad\times\delta(\mathcal{E}_j+\epsilon_n-\mathcal{E}_i-E^A_i)
(1-f(\epsilon_n)).
\end{align}
Here $\ket{\epsilon_{jn}}$ is a scattering state in the continuum part of the quasiparticle spectrum
and the factor $(1-f(\epsilon_n))$ gives the probability that this state is empty.
The matrix element in \eref{eq:3} is
the product of the off-diagonal matrix
element $s_{ji}$ times the overlap of the wavefunctions of bound and scattering states
at the junction,
\begin{equation}
\bra{j,\epsilon_{jn}}\hat H_{\rm nd}\ket{i,A_i}=-\tilde ts_{ji}\sum_{m}\psi_{jm}^{*}(\epsilon_n)
\sum_{n'}a_{in'},
\end{equation}
where $\psi_{jm}(\epsilon_n) = \bra{\epsilon_m} \epsilon_{jn} \rangle$ and
$\ket{\epsilon_m}$ are the eigenstates of $\hat H_{\qp +}$, see \eref{hqppm}.
Next we
calculate the wavefunctions for the continuum states by solving the scattering problem in the standard
$T$-matrix approach.\cite{Doniach-Sondheimer}

We focus again on the $\gamma_+$ subspace and write
$\hat H_{j+}=\hat H_{\qp +}+\hat H_{j1}$ with
$\hat H_{j1}=-\tilde ts_{jj}\sum_{nm,\sigma}\hat\gamma_{+n\sigma}^{\dagger}\hat\gamma_{+m\sigma}$
[cf. \eref{Hjpm}]. From
the Schr\"odinger equation, we have for the scattering states $\ket{\epsilon_{jn}}$
\begin{align}\label{eq:4}
\ket{\epsilon_{jn}}&=\ket{\epsilon_n}+\frac{1}{\epsilon_n-\hat H_{\qp +}+i0^+}\hat H_{1j}\ket{\epsilon_{jn}}
\nonumber\\
&=\left[\hat 1+\frac{1}{\epsilon_n-\hat H_{\qp +}+i0^+}\mathbb{T}_j(\epsilon_n)\right]\ket{\epsilon_n},
\end{align}
where
we have defined the T-matrix as
\begin{equation}
\mathbb{T}_{j}(\epsilon_n)=\hat H_{1j}+\hat H_{1j}[\epsilon_n-\hat H_{\qp +}+i0^+]^{-1}
\hat H_{1j}+\dots
\end{equation}
The T-matrix is related to the quasiparticle Green's function $G_j$ via
\begin{equation}
G_{j}=g+g\mathbb{T}_{j}g \, ,
\end{equation}
where $g$ is the (diagonal in momentum) bare quasiparticle Green's function
$g_n(\omega)=1/(\omega-\epsilon_n+i0^+)$. Using the inverse of this equation:
$\mathbb{T}_{j}=g^{-1}G_{j}g^{-1}-g^{-1}$, we find upon projecting
Eq.~(\ref{eq:4}) onto $\ket{\epsilon_m}$
\begin{equation}
\psi_{jm}(\epsilon_n)
=\lim_{\omega\rightarrow \epsilon_n}G_{j,mn}(\omega)(g_{n}(\omega))^{-1}.
\end{equation}
The Green's function, as obtained from the equations of
motion for $\gamma_{+n\sigma}$, is given by
\begin{align}\label{Gj}
G_{j,nm}(\omega)=\delta_{nm}g_{n}(\omega)-\frac{\tilde t
s_{jj}g_n(\omega)g_{m}(\omega)}{1+\tilde ts_{jj}\sum_p g_p(\omega)}.
\end{align}
Hence the continuum states are
\begin{equation}\label{eq:6}
\psi_{jm}(\epsilon_n)
=\delta_{nm}-\frac{\tilde ts_{jj}g_m(\epsilon_n)}
{1-i\pi\nu_0\tilde ts_{jj}\sqrt{\frac{2\Delta}{\epsilon_{n}-\Delta}}},
\end{equation}
where we have used that in the low energy limit $\sum_p g_{p}(\epsilon_{n})\approx
-i\pi\nu_0\sqrt{2\Delta/(\epsilon_{n}-\Delta)}$.

Using Eqs.~(\ref{eq:5}) and (\ref{eq:6}) we find
\begin{align}
\sum_{n}a_{in} &= 2^{1/4}\pi\nu_0\sqrt{\tilde t\Delta s_{ii}},\\
\sum_m \psi_{jm}^*(\epsilon_n)&=\frac{1}{1+i\pi\nu_0\tilde ts_{jj}\sqrt{2\Delta/(\epsilon_n-\Delta)}}\,.
\end{align}
Substituting these expressions into \eref{eq:3}, and considering explicitly the case of a phase qubit,
using the expressions for the matrix elements $s_{ij}$ given in \eref{sij_ph} finally
yields for the ionization rate of a single-channel junction
\begin{align}
\Gamma^A_j&=j\frac{{\omega_p}^2}{8\EZO}(1+\cos\varphi_0)
\frac{\sqrt{\frac{2\omega^A_j}{\mathcal{E}_j-\mathcal{E}_{j-1}-\omega^A_j}}}
{1+\frac{\omega^A_{j-1}}{\mathcal{E}_j-\mathcal{E}_{j-1}-\omega^A_j}}\nonumber\\
&\times
(1-f(\mathcal{E}_j-\mathcal{E}_{j-1}+E^A_j)),
\end{align}
with $\varphi_0$, $\omega_p$, and $\EZO$ defined in Sec.~\ref{sec:ph} and we used
that for a single-channel junction $E_J=\Delta (\pi \nu_0 \tilde{t})^2$.

The above result can be easily generalized to the case of $N_{ch}$
independent channels.
Assuming for simplicity identical transmission amplitudes,
the Andreev binding energy can be written as $\omega^A_j=2E_J{s_{jj}}^2/N_{ch}$ which for
a phase qubit reduces approximately to $\omega^A \approx E_J/N_{ch}$.
We assume $N_{ch}\gg 1$ sufficiently large so that $\omega^A \ll \EZO$
and obtain for the ionization rate of each occupied channel
\begin{align}
\Gamma^A_1&\approx
\frac{1}{4N_{ch}}\frac{{\omega_p}^2}{\EZO}
\sqrt{\frac{2\omega^A}{\EZO}}\,\frac{1+\cos\varphi_0}{2} \, .
\end{align}

A single ionization event is sufficient
to relax the qubit energy, and the probability of at least one ionization event taking place during time
$t$, when initially $N_{occ}\leq N_{ch}$ Andreev levels are
occupied, is given by $p=1-e^{-N_{occ}\Gamma^A_1 t}$. Introducing the total
ionization rate $\Gamma^A_{tot}=N_{occ}\Gamma^A_1$ and using \eref{ph_rel}, leads to the estimate
\be\label{ir_est}
T_1 \Gamma^A_{tot}\approx \frac{1}{4\sqrt{2\pi}}\frac{N_{occ}}{N_{ch}^{3/2}}
e^{\Delta/T_e}
\sqrt{\frac{E_J}{T_e}},
\ee
Defining the frequency shift $\delta\omega_q = \EZO - \omega_q$ and using \eref{om_q} to estimate its value,
we may eliminate $N_{occ}$ and rewrite the above as
\be
T_1\Gamma^A_{tot} \sim e^{\Delta/T_e}\sqrt{\frac{E_J}{N_{ch} T_e}} \, \frac{\delta\omega_q}{\EZO}
\sim 4\times 10^4 \frac{\delta\omega_q}{\EZO}
\ee
where we used $E_J/N_{ch} \sim 10^{-5}\Delta$ and $T_e \approx
140$~mK. Thus when $\delta\omega_q\gtrsim 10^{-4}\EZO$, the qubit
relaxation is likely dominated by the
ionization process, rather than by quasiparticle transitions within the continuum. However, we note
that the typical shift is much smaller than this,
$\delta\omega_q/\EZO \sim e^{-\Delta/T_e} \sim 3\times 10^{-7}$,
i.e. $p\approx 0.012$,
so the contribution of ionization to qubit relaxation is negligible unless $N_{occ}$ is anomalously large.

\subsection{Ionization by flux noise}
\label{app:fl_n}

As an example of an \textit{extrinsic} ionization mechanism, we consider here
low frequency ($\ll \EZO$) flux noise.
Small fluctuations $\delta \Phi_e(t) \ll \Phi_0$ of the external flux induce small fluctuations
$\varphi_1(t)$ of the phase difference $\varphi_0$,
\be
\varphi_1(t)= 2\pi\frac{\delta\Phi_e(t)}{\Phi_0}\frac{E_L}{E_L+E_J\cos\varphi_0}
\ee
[see \eref{ph_min}]. Since the low-frequency fluctuations do not induce qubit transitions, their
effect is accounted for by substituting $\varphi_0 \to \varphi_0+\varphi_1(t)$ into the diagonal matrix element
$s_{jj}$ in \eref{Hjpm}. At linear order in $\varphi_1$ we thus obtain the time-dependent perturbation
(in the $\gamma_+$ subspace)
\begin{equation}
\hat V(t)= -\tilde t\,\frac{\varphi_1(t)}{2}\cos\left(\frac{\varphi_0}{2}\right)
\sum_{n,m,\sigma}\gamma_{+n\sigma}^{\dagger}\gamma_{+m\sigma},
\end{equation}
Using Fermi's golden rule and
following similar steps as in the previous section, the total ionization rate
can be expressed as
\begin{align}
  \Gamma^A_{tot} &=
N_{occ}
  \left(\frac{E_J}{N_{ch}}\right)^{3/2}\left|\sin\frac{\varphi_0}{2}\right|\frac{1+\cos\varphi_0}{2}\\
&\times\int_{\omega_A}^{\infty}\!\!d\omega\,
  S_{\varphi\varphi}(\omega)\frac{\sqrt{\omega-\omega_A}}{\omega}
  (1-f(\omega+E_A))\nonumber.
\end{align}
where $S_{\varphi\varphi}(\omega)=1/(2\pi)\int e^{i\omega t}\langle\varphi_1(t)\varphi_1(0)\rangle dt$
is the phase fluctuation spectrum and the binding energy introduces a natural low-frequency cutoff.
For non-degenerate quasiparticles, $f(\omega_A+E_A)\ll 1$,
and a power-law spectrum of the form\cite{ithier,Yoshihara,Bialczak}
$S_{\varphi\varphi}(\omega)=(\delta\varphi)^2/(2\pi\omega^{\alpha})$, we obtain
\begin{align}
  \Gamma^A_{tot} &=\frac{(\delta\varphi)^2}{2\pi}
N_{occ}
  \left(\frac{E_J}{N_{ch}}\right)^{3/2}\left|\sin\frac{\varphi_0}{2}\right|\frac{1+\cos\varphi_0}{2}\nonumber\\
&\times\omega_A^{\frac{1}{2}-\alpha}\int_1^{\infty}dx\frac{\sqrt{x-1}}{x^{\alpha+1}}
\end{align}
For $\alpha=1$ (pure $1/f$ noise), the remaining integral is equal to
$\pi/2$ and since $\omega_A=2(E_J/N_{ch})\sin^2(\varphi_0/2)$, we arrive at
\begin{align}\label{eq:7}
  \Gamma^A_{tot} &=\frac{(\delta\varphi)^2}{4\sqrt{2}}
N_{occ}
  \left(\frac{E_J}{N_{ch}}\right)\frac{1+\cos\varphi_0}{2}.
\end{align}
The measured\cite{Yoshihara,Bialczak} magnitude of the fluctuations is small, $\delta\varphi \sim 10^{-6}$;
since
$E_J/N_{ch} \sim 10^{-5}\Delta$
and $N_{occ} \ll N_{ch} \lesssim 10^7$, we estimate this rate to be
much smaller than $1$~Hz.

\section{Modifications of the density of states}
\label{app:broad}

The logarithmic divergence of the daphasing rate and its regularization discussed in Sec.~\ref{sec:me}
are a consequence of the square root singularity of the BCS density of states at the gap edge. Here
we discuss two other mechanisms that also can regularize the divergence and show that
for Al-base qubits used at present they do not modify the results in the main text.

To begin with we consider the broadened density of states introduced by Dynes\cite{dynes}
to interpret experimental tunneling data. This phenomenological density of states is characterized by
a broadening parameter $\Gamma_D \ll \Delta$ and a finite density of subgap states.
These states give rise to an additional contribution to the dephasing
rate which we denote with $\Gamma_\phi^{sg}$; assuming quasi-equilibrium,
it is given by\cite{marth2}
\be\label{Gp_sg}
\Gamma_\phi^{sg}(T_e) = \frac{16E_J}{\pi} \left|A^d_s\right|^2 \left(\frac{\Gamma_D}{\Delta}\right)^2
\frac{T_e}{\Delta} \, .
\ee
and it is always smaller than the broadening, $\Gamma_\phi^{sg} \ll \Gamma_D$.
Comparing \esref{Gp_Te} and \rref{Gp_sg}, we see that a small broadening in the latter
can compensate for the exponential suppression of the quasiparticle occupation in the former. Then
we can distinguish three regimes: 1. at ``high'' temperatures, the dephasing rate is given
by \eref{Gp_Te}, since the broadening can be neglected in calculating $\Gamma_\phi$. The high-temperature
regime is defined by the condition $\Gamma_D \lesssim \Gamma_\phi(T_e)$; 2. at intermediate temperatures,
when $\Gamma_\phi^{sg}(T_e) \lesssim \Gamma_\phi(T_e) \lesssim \Gamma_D$,
the broadening of the density of states cannot be neglected. With logarithmic accuracy,
this amount to substitute $\Gamma_\phi \to \Gamma_D$ in the last term in \eref{Gp_scs} [and hence
replace the square bracket in \eref{Gp_Te} with $\ln (T_e/\Gamma_D)$; we note that since this substitution
affects only the logarithm, use of \eref{Gp_Te} still gives a correct order-of-magnitude estimate].
3. at low temperatures, such that $\Gamma_\phi(T_e) \lesssim \Gamma_\phi^{sg}(T_e)$ the subgap
contribution becomes dominant.

In recent measurements\cite{pekola} the intrinsic value of the broadening parameter in aluminum
was found to be small, $\Gamma_D/\Delta < 2\times 10^{-7}$. Using this value
and the results of the next section, our estimates show that the low-temperature regime is
entered for $T_e \lesssim 60$~mK. In experiments with superconducting resonators\cite{klap}
as well as qubits\cite{paik,corcoles} the quasiparticle effective temperature
is larger, $T_e \sim 140$~mK, so we can neglect the subgap contribution to the dephasing rate
for Al-based qubits, which we focus on in this paper.
However, the subgap contribution may be relevant in other systems, such as qubits fabricated with
niobium.\cite{marth2}

While the above consideration are based on a phenomenological model, an \textit{intrinsic} modification
of the continuum part of the  density of states near the junction is due to the presence of Andreev bound
states. They modify the square root singularity into a square root threshold,
\be\label{dos_tr}
\sqrt{\frac{2\Delta}{\omega -\Delta}} \to \frac{\sqrt{2\Delta}\sqrt{\omega-\Delta}}{\omega - E^A}
\ee
with $E^A$ the energy of the bound state defined in \eref{as_be} (here we consider for simplicity
the single channel case). The above substitution can be obtained
using \eref{Gj} for the Green's function to calculate the density of states. Assuming the binding energy
$\omega^A = \Delta - E^A$
to be small compared to the typical quasiparticle energy, $\omega^A \ll \delta E$, we find that
the substitution \rref{dos_tr} would lead to the replacement of $\Gamma_\phi$ with $\omega^A$ in
the right hand side of \eref{Gp_scs}. In quasi-equilibrium this amount to replacing
the square brackets in \eref{Gp_Te} with
\be\label{be_sub}
\ln \frac{T_e}{\omega^A} \sim \ln \frac{T_e}{E_J} + \ln N_{ch}
\ee
where $N_{ch} \gg 1$ is the number of channels in the junction. We note that the tunneling
limit we are considering consists in taking the transmission amplitude $\tilde{t} \to 0$ at
finite $E_J$, which implies $N_{ch} \to \infty$. Then in this limit the self-consistent approach
is justified with logarithmic accuracy as explained in Appendix~\ref{app:scd}.
.


\begin{thebibliography}{30}
\bibitem{ch_ind}We will reintroduce the channel index where necessary, e.g. in Sec.~\ref{sec:and}
\bibitem{f0est}Here and in subsequent formulas we take the distribution function to be a smooth
function of energy over an interval of order $\delta E$ above the gap and to vanish quickly at higher
energies.
\bibitem{phq_app}The main approximation used here is the harmonic one for the potential energy of the
phase variable in \eref{Hq0}, justified by the smallness of the
fluctuations of the phase around a minimum. However, in \eref{ph_me} the dependance on $(E_C/\EZO)^2$ is
due to the cancellation of the leading terms in the difference $s_{11}-s_{00}$ [cf. \eref{Asdef}].
Then a more careful analysis shows that corrections to the wavefunctions due to the cubic anharmonicity of
the potentail should be considered when evaluating $A^d_s$. Since the corresponding
corrections are of order unity for typical phase qubit parameters, they do not affect our conclusions in
the text after \eref{ph_Tecond}.
\bibitem{xAnote}This exponential factor coincides with the average occupation
probability $f(\Delta)$ at the gap edge; in Refs.~\onlinecite{prl,prb} two contributions to the frequency, one
coming from the Andreev states and the other from the bulk quasiparticles with energy at the gap edge,
were combined together since for tunnel junctions $x_\qp^A = f(\Delta)$.
More generally, the two contributions must be considered separately. For example, for junctions with highly
transparent channels we have $x_\qp^A = f(E_A) \neq f(\Delta)$, and when considering fluctuations of
the occupations their relative magnitudes scale differently for bound and free quasiparticles
(as $1/\sqrt{N}$ and inverse square root of the volume, respectively).
\bibitem{hma}More precisely, we here perform the first step of the Markov approximation -- the subsitution
$\rho_+(\tau) \to e^{-i\EZO(t-\tau)}\rho_+(t)$. The second step (the extension of the lower
integration limit) will be taken at the end of the calculation.
\bibitem{fl_perf}We note that in the flux qubit the inductor is assumed to be lossless, while in
the fluxonium the effective inductor formed by the junction array has finite losses.
\end{thebibliography}
\end{document}